# Genome-wide Causation Studies of Complex Diseases

Rong Jiao[1], Xiangning Chen[2], Eric Boerwinkle[3] & Momiao Xiong[1*]

[1]Department of Biostatistics and Data Science, School of Public Health, The University of Texas Health Science Center at Houston, Houston, Texas, USA

[2] Nevada Institute of Personalized Medicine, University of Nevada, Las Vegas, Nevada, USA

[3]Epidemiology, Human Genetics & Environmental Sciences, School of Public Health, University of Texas Health Science Center at Houston, Houston, Texas, USA

[*]Address for correspondence and reprints: Dr. Momiao Xiong, Department of Biostatistics and Data Science, School of Public Health, The University of Texas Health Science Center at Houston, P.O. Box 20186, Houston, Texas 77225, (Phone): 713-500-9894, (Fax): 713-500-0900, E-mail: Momiao.Xiong@uth.tmc.edu.

**ABSTRACT**

Despite significant progress in dissecting the genetic architecture of complex diseases by genome-wide association studies (GWAS), the signals identified by association analysis may not have specific pathological relevance to diseases so that a large fraction of disease causing genetic variants is still hidden. Association is used to measure dependence between two variables or two sets of variables. Genome-wide association studies test association between a disease and SNPs (or other genetic variants) across the genome. Association analysis may detect superficial patterns between disease and genetic variants. Association signals provide limited information on the causal mechanism of diseases. The use of association analysis as a major analytical platform for genetic studies of complex diseases is a key issue that hampers discovery of the mechanism of diseases, calling into question the ability of GWAS to identify loci underlying diseases. It is time to move beyond association analysis toward techniques enabling the discovery of the underlying causal genetic strctures of complex diseases. To achieve this, we propose a concept of a genome-wide causation studies (GWCS) as an alternative to GWAS and develop additive noise models (ANMs) for genetic causation analysis. Type I error rates and power of the ANMs to test for causation are presented. We conduct GWCS of schizophrenia. Both simulation and real data analysis show that the proportion of the overlapped association and causation signals is small. Thus, we hope that our analysis will stimulate discussion of GWAS and GWCS.

## INTRODUCTION

Although significant progress in dissecting the genetic architecture of complex diseases by GWAS has been made, the overall contribution of the new identified genetic variants is small and a large fraction of causal genetic variants is still hidden. Association is to measure dependence between two variables or two sets of variables in the data and to use these dependencies for prediction that is not dealing with causal problems (Altman and Krzywinski, 2015; Lopez-Paz 2016). Association analysis may detect superficial patterns between disease and genetic variants. Its signals provide limited information on the causal mechanism of diseases (Kahrilas and Kahrilas 2019). Association analysis has been a major paradigm of genetic analsyis of complex diseases for almost a century. Understanding the etiology and mechanism of complex diseases using association analysis remains elusive. Most genetic questions to uncover the mechanism of the disease is causal in nature. Causation analysis is an essential to the genetic analysis of complex diseases, yet ignored for long time (Lopez-Paz 2016; Kreif and DiazOrdaz 2019).

It is well recognized that association analysis is not a direct method to discover the causal mechanism of complex diseases. Many investigators think that “association is essential to causation” and hope that we can successfully shift from association to causation (Jones et al. 2017). A current paradigm to make transition from association to causation is through omics analysis (Clyde 2017; Ongen et al. 2017). However, such approaches have two limitations. First most current omics analysis still detect association signals. For example, eQTL analysis that tests for association of a discrete variable (genetic variant) with a continuous variable (gene expression) is still association analysis. Observed association may not lead to inferring a causal relationship ( Orho-Melander 2015; Lee et al. 2018). The recent study has found that “association signals tend to be spread across most of the genome (Boyle et al. 2017). The paradigm of GWAS with eQTL may still fail to identify the causal paths from genetic variants to disease. Second, the lack of association may not be necessary to imply the absence of a causal relationship (Callaway et al. 2017). The set of causal loci including causal QTL, causal eQTL, and causal mQTL is not the subset of association loci that are identified in QTL, eQTL and mQTL analysis simply because QTL, eQTL and mQTL analysis are based on regression. A large proportion of causal loci may not be discovered by association analysis. Finding causal SNPs only via searching the set of associated SNPs may miss many causal SNPs. In summary, the use of association analysis as a major analytical platform for genetic studies of complex diseases is a key issue that hampers identification of causal SNPs and discovery of the causal mechanisms of the diseases.

Distuinguishing causation from association is an age-old problem. Methods for causation analysis that is one of the greatest challenging problems in science and technology need to be developed as an alternative to association analysis (Zenil et al. 2019). Without a proper causal analysis, to fully detect causal SNPs is not possible in general. Intutively, causation implies that changes in one variable will directly make changes in the other (Jaffe 2010). The essential distinction between association and causation relies on what the response will be if we intervene in the system (Lattimore and Ong 2018). There are two types of causal inference: intenventional causal inference and observational causal inference (Kaplan 2018). Interventional causal inference learns the effect of taking an action directly via experiments, for example, randomized controlled trials. Interventional experiments are a gold standard for causal inference. However, since in human genetics, we cannot change the genetic materials of human subjects, experimental interventions are unethical and infeasible. Therefore, it is essential to develop statistical methods and algorithms to predict the outcomes of an intenvention from passive observation (Spirtes et al. 2000; Lattimore and Ong 2018). In this paper, we take an observational causal inference approach to identifying causal SNPs.

Although we infer causation from observation data, our concept of causation is derived from intervention (Pearl 2019). In principle, causal inference is based on interventional distribution. The do-calculus is used as an essential concept for causal inference, which can simplify the expression for an interventional distribution. Repeated applications of the do-calculus will lead to an expression containing only observational quantities that can be used to estimate the interventional distribution from observational data (Lattimore and Ong 2018). Therefore, the do-operation is a key concept that makes observational causal inference feasible.

Three essential frameworks: causal Beyesian networks, structural equation models and counterfactuals have been developed for observational causal inference (Rosenbaum and Rubin 1983; Pearl 2000; Peters et al. 2014; Peters et al. 2017; Xiong 2018; Lattimore and Ong 2018). In history, causal Beyesian networks, structural equation models and counterfactuals developed relatively independently in different fields, but they can be unified using interventional queries with do-calculus (Lattimore and Ong 2018). This allows methods and algorithms developed within one framework to be easily applied to one another, and also allows predictions about the consequences of intervening upon (rather than merely observing) variables, and provides a method of evaluating counterfactual claims. Therefore, we will use do-calculus as an unified framework for causal inference.

Similar to GWAS which investigates the dependence relationship between SNPs and disease at a time, GWCS investigates the causal relationship between SNPs and disease at a time, referred to as bivariate causal discovery. The traditional causal inference theory infers causal relationships among more than three variables and cannot be applied to bivariate causal discovery. Only recently the independence of cause and mechanism (ICM) and functional causal model, specifically additive noise models (ANMs) (Peters et al. 2017; Xiong 2018), has been proposed. ICM and discrete ANMs can be applied to GWCS.

For a long time, many genetic epidemiologists hold a view that that causal inference from observational data is impossible. Some views and concepts that misunderstand causation widely exist in genetic epidemiology. There is also lack of algorithms for causal inference in genetics. Purpose of this paper is to rigorously define causation, clarify concept of causation and association and develop effective causal models and algorithms that can be easily used to discover causal structure in genetic analysis. While there is increasing evidence that association signals provide limited information on causes of disease and some investigators call the future of the GWAS into question (Callaway 2017), the modern causal inference theory provides powerful tools for bivariate causal discovery. In the past two dacades, causal theory has been well developed and is becoming an important component of artificial intelligence (AI). "Reasoning in causal terms is omnipresent, from fundamental physics to medicine, social sciences and economics, and in everyday life" (Barrett et al. 2019).

It is urgent to develop concepts and theory to show that under the right conditions and assumptions, causal-effect relationship between two variables can be inferred from purely observational data. It is time to develop a new generation of genetic analysis to shift the current paradigm of genetic analysis from association analysis to causal inference. To make the shift feasible, we will rigorously use do-calculus to model intervations, define the concept of causation, unify counterfactuals, functional causal models and ICM, and investigate the connections and difference between association and causation. ANMs are easily used causal models. We use ANMs $Y = f_Y(X) + N_Y$ where $Y$ represents the disease status, $X$ represents the indicator variable for the genotype of a SNP, and $N_Y$ represents some residual term, as a general framework to distinguish causal directions and develop a new ANM-based statistic to test causation of SNP locus with the disease, which will be used for GWCS of complex disease. Under the assumption of no counfounders (causal model with counfounders will be discussed somewhere else), we investigate the identifiability of the ANM-based statistics for bivariate (SNP and disease) causal discovery. Since an analytical form for the distribution of the causal test statistic is difficult to derive, permutation methods will be used to compute the distribution of the causal test.

To evaluate its performance for genetic causal analysis, we use large scale simulations to calculate the type I error rates of of the ANM-based statistics to test causation and to ompute its power under various conditions. To further evaluate its performance, an ANM-based casual test is applied. The proposed method is applied to the CATIE-MGS-SWD schizophrenia (SCZ) study dataset with 8,421,111 common SNPs typed in 13,557 individuals to perform GWCS of SCZ. To further investigate the properties of the ANM-based causal test, we will investigate the prediction ability of causal SNPs and impact of linkage disequilibrium (LD) on the causation analysis. Our purpose is to provide a detailed analysis of GWAS and GWCS as a response to the comments about the ability of GWCS to identify disease causing loci (Orho-Melander 2015).

**Basic concepts of association and causation**

In this section, we briefly introduce causal inference theory to make this section as self-contained as possible. We assume that two variables $x$ and $y$ are considered. Their joint distribution is denoted by $P(x, y)$. Association between two variables $x$ and $y$ are defined as dependence between them. Statistically, association between $x$ and $y$ is defined as

$$P(y|x) \neq P(y). \quad (1)$$

Statistical dependence is a symmetric concept: if the variable $x$ to depends on the variable $y$, then the variable $y$ also depends on the variable $y$.

Classical machine learning and statistical methods, built on pattern recognition and association analyses, are insufficient for causal reasoning. The science of causal reasoning is developing in various disciplines. In different disciplines, there may be different definitions of causations. Four key approaches have emerged: structural equation models, causal Bayesian networks, counterfactuals and independence of cause and mechanism (ICM) (Lattimore and Ongv 2018; Marsala 2015; Peters et al. 2017; Xiong 2018). The four schools of causality have been recently unified. Intervention calculus (do-calculus) can be taken as an unifuing language for causal inference.

*Intervention calculus*

The purpose of intervention calculus is to describe the mathematical conditions under which we can make causal inference from observational data. Intuitively, causation is defined as the encoding of potential outcomes under intervention. Intervention is surgeries on mechanism. In other words, changes in one variable under intervention will affect the outcomes of another variable and hence can be used to measure effects of intervention (action).

We consider two variables $X$ and $Y$. A causal model can be defined by intervention (action) as follows. If we $do\, X$ (forcing the random variable $X$ to take a specified value), then $Y$ will be affected. Causation analysis investigates prediction of the effects of actions that perturb the observed system (Mooij et al. 2016).

We use $P(Y|do\ (X))$ to denote the distribution of $Y$ conditional on an intervention that sets $X = x$. Now $X$ causing $Y$ can be methatically defined as

$$P(Y|do(X_1)) \neq P(Y|do(X_2)) \text{ for some } X_1, X_2, X_1 \neq X_2. \tag{2}$$

If $X$ causes $Y$ $(X \rightarrow Y)$, then in general, we have

$$P(Y|X) = P(Y|do(X)) \neq P(Y). \tag{3}$$

However, $P\big(X\big|do\ (Y)\big) = P(X) \neq P(X|Y)$, or if $Y$ causes $X$ $(Y \rightarrow X)$ then

$$P\big(Y\big|do(X)\big) = P(Y) \neq P(Y|X).$$

Although the joint probability can be factorized in terms of marginal distribution and conditional istribution as

$$P(XY) = P(X)P(Y|X) = P(Y)P(X|Y),$$

If $X$ causes $Y$ $(X \rightarrow Y)$, we have the factorization: $P(XY) = P(X)P(Y|do(X))$, but in this case $(X \rightarrow Y)$, we do not have $P(XY) = P(Y)P(X|do(Y))$, i.e., $P(XY) \neq P(Y)P(X|do(Y))$, the joint probability of $X$ and $Y$ cannot be factorized in terms of marginal distriation $P(Y)$ and interventional probability distribution $P(X|do(Y))$ unless $X$ and $Y$ are independent.

For the genetic problem, $Y$ represents a disease status and $X$ represents a genotype. The action do $X$ means that changing genotype $X$ is conducted (for human subject this is impossible, but for animal, it can be done by genome editing). Intervention calculus implies that if $X$ causes disease, then $P(Y|doX)) = P(Y|X)$, otherwise if $X$ is not disease lcous, then $P(Y|do(X)) = P(Y) \neq P(Y|X)$.

Do-calculus can also be defined as $E[Y|do(X)]$. If effect variable $Y$ is a binary variable, then we have

$$E[Y|do(X)] = P(Y = 1|do(X)).$$

The various relationships between marginal, conditional and interventional distributions of $X$ and $Y$ under causation and association are summarized in Figure 1. Figure 1 (d) clearly demonstrates differences between association and causation. Although temperate in the room and thermometer are associated, since temperature causes changes in thermometer, change in thermometer cannot change the temperature in the room, i.e., $P(X|do(Y)) = p(X)$.

In summary, association is studied by observed conditional distribution and causation is investigated by interventional distribution where causal effect is determined by the effect of hypothetic manipulation of an input on an output. In other words, association is investigated by seeing and causation is investigated by doing. To illustrate the the difference between seeing and doing, we present the following example:

**Example 1**

Consider

$$Z_i \sim N(0,1),$$

$$X_i \leftarrow 2Z_i,$$

$$Y_i \leftarrow 2X_i + Z_i^2.$$

Then, from the observed data generated by this model, we can estimate that

$$E[Y|X = 1] \approx 2\frac{1}{4}.$$

Next we perform intervention $do\ (X = 1)$ . Then, the intervened generative model is

$$Z_i \sim N(0,1),$$

$$X_i \leftarrow 1,$$

$$Y_i \leftarrow 2X_i + Z_i^2.$$

Then, we obtain that

$$E[Y|do(X = 1)] \approx 3\ .$$

This clearly demonstrates that seeing and doing are quite different.

*Conuterfactual*

Causality can also be defined in terms of potential outcomes or counterfactuals in the Neyman-Rubin causal model (Rosenbaum and Rubin 1983). We can read interventional distribution $P\big(Y\big|do(X)\big)$ as counterfactual questions: "what would have been the distrution of $Y$ had $X = x$?" (Lopez-Paz 2016). Intuitively, counterfactuals assume the presence of an alternative world where everything is the same as the factual world, except for the alternative (hypothetical) intervention and its effects.

For simplicity, consider a binary treatment $T = \{0,1\}$ or potential intervention, where $T = 1$ indicates the treatment (intervention) and $T = 0$ indicates the control (no intervention). Each individual has two potential outcomes, $\{Y_i^1, Y_i^0\}$, one for each value of the treatment:

$Y_i^1$: potential outcome if the individual received a treatment ($T_i = 1$) ,

$Y_i^0$: potential outcome if the individual received no treatment ($T_i = 0$) .

This implies that a potential outcome is the outcome that would be realized if the individual received a specific value of the treatment (intervention). A SNP has two alleles. We can define

$$T = \begin{cases} 1 & \text{allele } A \\ 0 & \text{allele } a \end{cases} .$$

The potential outcome is 1 if the individual is affected, potential outcome is 0 if the individual is normal. Let $X$ be the set of contexts (covariates).

The "fundamental problem of causal inference" (Holland 1986) is that we can only observe one of the potential outcomes rather than both of them. The unobserved (missed) potential outcome is called "counterfactual" outcome. Similar to do-calculus, coundetrfactual can be defined as stating that $\bar{Y}$ would change to $Y$ if it were $T$. In other words, we imagine that value $Y$ would be taken if we did hypothetical intervention $T$. Causal effects are defined as differences in counterfactual variables. In other words, it measures difference between what would have happened if we did one thing versus what would have happened if alternatively, we did something else (Lattimore and Ong 2018). A brief overview about counderfactual theory is summarized in Supplementary A.

*Structural equation model and independence of cause and mechanism (ICM)*

The third language of causation which we inrtoduce is structural equation moels (SEMs). SEMs can be used to model causal relationships between some given variables, where each variable is expressed as a function of some other variables (its causes or treatments) as well as some noise (Nowzohour and Bühlmann 2016, Xiong 2018). The model consists of three essential components: (1) causal structure, (2) the functional dependence among causal and effect variables, and (3) the joint distribution of the noises. We assume that (1) there are no unobserved variables and hence that the noise terms are independent and (2) the difference between the effect variable and some noise term is a deterministic function of the causal variables. In this paper, we focus on bivariable causal discovery. The SEMs for two variables is defined as (Lattimore and Ongv 2018)

$$X = f_x(\varepsilon_x),\ Y = f_y(X, \varepsilon_Y),\quad \varepsilon_x \perp\!\!\!\perp \varepsilon_Y\ , \qquad (4)$$

where $\varepsilon_x$ and $\varepsilon_y$ are noises or exogenous random variables. If the functions $f_x$ and $f_y$ are free form, the SEMs are called nonparametric structural equation models. One can

The structural equation model (4) encodes the assumption that the outcome $Y_i$ for an individual $i$ is caused by the cause (treatment) $X_i$ which the individual receives and other factors $\varepsilon_y$ that are indepdent of cause $X$. The SEMs describe the causal effects of performing real-world interventions or experiments on their variables $X$.

Although conditional independences can be used to make causal inference from the data under study, conditional independence cannot be applied to causal analysis under two variaables (Lopez-Paz 2016). Consider observational causal inference for two random variables, $X$ and $Y$. We want to infer whether $X \rightarrow Y$ or $X \leftarrow Y$. Unfortunately, the absence of a third random variable prevents us from measuring conditional independences. To overcome this limitation, in the past decade, observational causal inference methods that are not based on conditional independence have been developed. One of them is the widely used Independence of cause and mechanism (ICM) principle.

Independence of Cause and Mechanism (ICM) assumes that causes and mechanisms are chosen independently by nature is a recently proposed principle for causal reasoning and causal learning (Janzing and Sch¨olkopf 2010; Shajarisales et al. 2015; Peters et al. 2017). ICM assumes that the mechanism that generates effect from its case contains no imformation about the the cause. Assume that $X$ is a cause and $Y$ is an effect. The joint distribution $P(X,Y)$ can be decomposed into

$$P(X,Y) = P(X)P(Y|do\,(X)). \tag{5}$$

The conditional distribution $P(Y|do(X))$ is a mechanism that generates effect $Y$ from cause $X$. The conditional distribution $P(Y|do(X))$ is independent of the distribution of cause $P(X)$. If $X$ causes $Y$ then $P(Y|do(X)) = P(Y|X)$ . The conditional distribution $P(Y|X)$ contains no information about marinal distribution of cause $P(X)$. Therefore, the ICM postulates that the conditional distribution of each variable given its causes contains no information about its cause.

SEMs, ICM and counterfactuals developed relatively independently in different fields. However, it is shown that they can be unified under some assumptions using interventional queries with do-calculus (Supplementary B). This allows methods and algorithms developed within one framework to be easily applied to one another and provides foundation for interpretation of ANMs and justification of GWCS.

**Additive noise models**

In the previous section, we showed that the SEMs, the ICM and counterfactual approach to causal inference are equivalent in general. To facilitate the application of causal inference to the real world, we need simpler methods to implement these general approaches to causal inference. In this paper, we propose to use discrete additive noise models (ANMs) that are based on the ICM principle, as a tool for GWCS. We assume that there is no confoundeing, no selection bias and no feedback between the cause and effects (Mooij et al. 2016). The methods for causality analysis with confoundeing will be presented else where.

Let $Y$ be a binary variable to indicate disease status: $Y = 1$, presence of disease and $Y = 0$, normal and $X$ be a genotype indicator variable:

$$X = \begin{cases} 0 & dd \\ 1 & Dd \\ 2 & DD \end{cases},$$

where $D$ is a disease allele.

Let $m$ be an integer. Assume that $Z = km + r, r = 0, 1, \ldots, m-1$. $Z$ is called a m-cyclic random variable, if $Z$ takes the remainder $r$ as its value. Now we define a discrete ANMs for genetic causation analysis.

Let $X$ and $Y$ be 3 and 2-cyclic random variables, respectively. An ANM from $X$ to $Y$ is defined as (Peters et al. 2011)

$$Y = f_y(X) + N_y, \quad X \perp\!\!\!\perp N_y, \tag{6}$$

where $f_x$ is an integer function and $N_y$ is a 2-cyclic noise variable.

An ANM is called reversible if there is also an ANM:

$$X = f_x(Y) + N_x, \quad Y \perp\!\!\!\perp N_x, \tag{7}$$

where $N_x$ is a 3-cyclic noise variable.

In practice, there may be multipe potential cuasations $X_2, \dots, X_k$. However, only one causation $x_1$ is explicitely considered in the model equation (6). Other cuasations $X_2, \dots, X_k$ are unobserved. Their causal effects to $Y$ are accounted for by residual. Then, we can show that the following model

$$Y = f_Y(X_1) + \tilde{N}_Y,\ \ X_1 \perp\!\!\!\perp \tilde{N}_Y \tag{8}$$

where the effects of $X_2, \dots, X_k$ on $Y$ are included in $\tilde{N}_Y$ still holds if we assume that $X_1 \perp\!\!\!\perp X_2, \dots, X_1 \perp\!\!\!\perp X_K$. Its extension to multiple dependent causations is more complicated and will be presented elsewere.

It is well known that the set of joint distributions $P(X, Y)$ that allow the ANM in both forward and backward directions is very small. In other words, in general, the direction of the ANMs is identifiable (Peters et l. 2011). Assumptions for identificability of the direction of the ANMs are summarized in Supplementary C.

In our cases, $X$ is an indicator variable for genotypes and $Y$ is a binary variable for disease status. In Supplementary C, we show that in general, reversible is impossible and hence the direction of the ANMs is identifiable.

**Numerical algorithms to implement ANMs for genetic causal analysis**

To implement the ANMs to identify a causal SNP, we used the numerical algorithm that was presented in the paper (Peters et al. 2011) to test the causal relathinsip between the SNP and disease. The algorithm is summarized as follows (Hu et al. 2018).

**Algorithm to implement the discret ANMs for genetic causal analysis:**

Assume that qualitative trait data $Y$ and indicator variable $X$ for the genotypes of a SNP are available.

1. To infer direction $X \to Y$, we regress the trait $Y$ on the genotype indicator variable $X$: $Y = f(X) + N_Y$. Calculate the residuals $\widehat{N}_Y = Y - \hat{f}(X)$.
2. To infer potential causal direction $Y \to X$, we fit the following nonlinear integer regression to the data: $X = g(Y) + N_X$. Calculate the residuals $\widehat{N}_X = X - \hat{g}(Y)$ .
3. Test for independence between the residuals and potential causation. If $\widehat{N}_Y$ and $X$ are independent ($\widehat{N}_Y \perp\!\!\!\perp X$) , and $\widehat{N}_X$ and $Y$ are not independent, then $X$ causes $Y$ ($X \to Y$)

   If both $\widehat{N}_Y$ and $X$, and $\widehat{N}_X$ and $Y$ are not independent or if both $\widehat{N_Y}$ and $X$, and $\widehat{N}_X$ and $Y$ are independent, then no causation conclusion can be made.

Nonlinear integer regressions to implement the ANMs have two important features. First, in general, we do not have general functional forms for nonlinear integer functions. We usually, investigate all possible mapping (functions) from $X$ to $Y$ and evaluate their values of lost function. Second, the ordinary regression usually minimizes the sum of square of errors. However, in the above algorithm, in addition to evaluate the loss function, we still need to test the independence between the the regressor and residuals. Therefore, Peters et al. (2011) suggested using a dependence measure (DM) between regressor and residuals as a lost function. We adopt the discrete regression with dependence measure minimization procedure for genetic causal analysis (Peters et al. 2011).

**Discrete nonlinear regression with dependence measure minimization for genetic causal analysis**

**Step 1:** Calculate the sampling distribution $\hat{P}(X, Y)$.

**Step 2:** Initialization.

$$f^{(0)}(x^i) = argmax_y \hat{P}(X = x^i, Y = y), t = 0.$$

**Step 3: Repeat**

$t = t + 1;$

**Step 4: for** $i = 1, \ldots, n$ do

**Step 5:** $f^{(t)}(x^i) = argmin_Y DM(X, Y - f^{(t-1)}_{x_i \to y}(X))$

**end for**

**Step 6: until** $||f^{(t)} - f^{(t-1)}||_2 < \varepsilon$ or $\left(\hat{N}_Y = Y - f^{(t)}(X)\right) \perp\!\!\!\perp X$ , or $t = T$, where $\varepsilon$ and $T$ are pre-specified.

A $\chi^2$ test statistic will be used as the dependence measure (DM). Specifically, we formulate a $2 \times 3$ contingence table (Table 1). In the ANM equation (6), $N_y$ is assumed as a 2-cyclic noise variable. Let $n_0$ and $n_1$ be number of individuals with $N_Y = 0$ and $N_Y = 1$, respectively. Let $n = n_0 + n_1$. Consider three genotypes: $dd, dD$ and $DD$. Let $N_Y = 0$, $a_{11}, a_{12}$ and $a_{13}$ be the number of individuals with genotypes $dd, dD$ and $DD$, respectively. Let $N_Y = 1$, and $b_{11}, b_{12}$ and $b_{13}$ be the number of individuals with genotypes $dd, dD$ and $DD$, respectively. Define the marginal frequencies as shown in Table 1. Then, we obtain

$$E[a_{1j}] = \frac{n_0(a_{1j}+b_{1j})}{n}, j = 1,2,3, \text{ and } E[b_{1j}] = \frac{n_1(a_{1j}+b_{1j})}{n}, j = 1,2,3.$$

Then, the test statistic for testing independence is defined as

$$DM = \sum_{j=1}^{3} \left[\frac{(a_{1j}-E[a_{1j}])^2}{E[a_{1j}]} + \frac{(b_{1j}-E[b_{1j}])^2}{E[b_{1j}]}\right]. \quad (9)$$

Under the null hypothesis of independence, the test statistic $DM$ is distributed as a central $\chi^2_{(2)}$ distribution with 2 degrees of freedom. If SNPs involve rare variants, the expected counts of many cells will be small. Fisher's exact test should be used to test for independence.

The statement that there is no causal relationship between the SNP and disease implies that neither causations $X \to Y$ nor causation $Y \to X$ holds. Let $DM_{X \to Y}$ and $DM_{Y \to X}$ be the $\chi^2$ statistics for testing causations $X \to Y$ and causation $Y \to X$, respectively.

The null hypothesis for testing causal relationships between two random variables $X$ and $Y$ is

$H_0$ : no causation between two random variables $X$ and $Y$ .

The statistic to test the causal relationsips between two randoom variables $X$ and $Y$ is defined as

$$T_C = |DM_{X \to Y} - DM_{Y \to X}| \; . \tag{10}$$

When $T_C$ is large, either $DM_{X \to Y} > DM_{Y \to X}$ which implies $X$ causes $Y$, or $DM_{Y \to X} > DM_{X \to Y}$ which implies that $Y$ causes $X$. When $T_C \approx 0$, this indicates that no causal decision can be made. Since $DM_{X \to Y}$ and $DM_{Y \to X}$ may be dependent, a closed, analytic expression for the distribution of $T_C$ is not yet known (Bausch 2012). Although a computational algorithm to numerically calculate the distribution of $T_C$ is available, in this paper we will use the permulation test to calculate the P-value of the test $T_C$.

**Distance Correlation as a Causation Measure**

In previous sections, we introduce the basis principal for assessing causation $X \to Y$ that the distribution $P(X)$ of causal $X$ is independent of the causal mechanism or conditional distribution $P(Y \mid X)$ of the effect $Y$, given causal $X$. Now the question is how to assess their independence. The Pearson correlation coefficient $\rho(X, Y)$, the widely-used classical meaure of dependence measures linear dependence between two random variables $X$ and $Y$, and in the bivariate normal case, $\rho(X,Y)=0$ is equivalent to independence between $X$ and $Y$. If the distributions of $X$ and $Y$ are not normal, then $\rho(X,Y)=0$ may not imply independence between $X$ and $Y$. Recently distance correlation that can be applied to all distributions with finite first moments is proposed to measure dependence between random vectors which allows for both linear and nonlinear dependence (Sze´kely et al. 2007, 2009). Distance correlation extends the traditional Pearson correlation in two remarkable directions:

(1) Distance correlation extends the Pearson correlation defined between two random variables to the correlation between two sets of variables with arbitrary numbers;

(2) Zero of distance correlation indicates independence of two random vectors.

Consider two vectors of random variables: $p$ - dimensional vector $X$ and $q$ - dimensional vector $Y$. Let $P(X)$ and $P(Y)$ be density functions of the vectors $X$ and $Y$, respectively. Let $P(X,Y)$ be the joint density function of $X$ and $Y$. There are two ways to define independence between two vectors of variables: i) density function definition and ii) characteristic function definition. In other words, if $X$ and $Y$ are independent then either

$$\text{i)}\ P(X,Y)=P(X)P(Y)$$

or

$$\text{ii)}\ f_{X,Y}(t,s)=f_X(t)f_Y(s),$$

where $f_{X,Y}(t,s)=E[e^{i(t^Tx+s^Ty)}]$, $f_X(t)=E[e^{it^Tx}]$ and $f_Y(s)=E[e^{is^Ty}]$ are the characteristic functions of $(X,Y)$, $X$ and $Y$, respectively. Therefore, we can use both distances $||P(X,Y)-P(X)P(Y)||$ and $\| f_{X,Y}(t,s)-f_X(t)f_Y(s)\|$ to measure the dependence between two vectors $X$ and $Y$. Since characteristic function $f$ is a complex-valued function, its norm is defined as $|f|^2=f\bar{f}$. Definition of The distance covariance $dCov^2(X,Y)$ between two random vectors $X$ and , distance variance $dVar^2(X)$ , and algorithms for their calculations are briefly introduced in Supplementary D. Square of correlation $R^2(X,Y)$ is defined as

$$R^2(X,Y) = \begin{cases} \frac{dCov^2(X,Y)}{\sqrt{dVar^2(X)dVar^2(Y)}} & dVar^2(X)dVar^2(Y) > 0 \\ 0 & dVar^2(X)dVar^2(Y) = 0 \end{cases}. \quad (11)$$

Now we propose to use distance correlation to measure the dependence between the distributions $P(X)$ and $P(Y|X)$. Assume that $X$ takes $m$ different values and $Y$ takes $\widetilde{m}$ different values. Define two vectors $P(X) = [P(X_1), \dots, P(X_m)]^T$ and

$$P(Y|X) = [P(Y_1|X_1), \dots, P(Y_{\widetilde{m}}|X_1), \dots . P(Y_1|X_m), \dots, P(Y_{\widetilde{m}}|X_m)]^T.$$

A meaures for causal directions $X \to Y$ and $Y \to X$ are defined as

$$C_{X \to Y} = 1 - R(X, P(Y|X)) \quad (12)$$

and

$$C_{Y \to X} = 1 - R(Y, P(X|Y)) , \quad (13)$$

respectively.

A measure for quantifying the strength of causal relationships between $X$ (genetic variant) and $Y$ (disease phenotype) can be defined by

$$CR = |C_{X \to Y} - C_{Y \to X}|. \quad (14)$$

Using Theorem 3 in the paper (Sze´kely et al. 2009) , we can show that

$$0 \le C_{X \to Y} \le 1 \text{ and } C_{X \to Y} = 1 \text{ if and only if } X \to Y .$$

Similarly, we can show that

$$0 \le C_{Y \to X} \le 1 \text{ and } C_{Y \to X} = 1 \text{ if and only if } Y \to X.$$

Consider linear transformations $U = a_1 + b_1 D_1 X$ and $V = a_2 + b_2 D_2 Y$ where $D_1$ and $D_2$ are orthonormal matrices, then we can show that $C_{U \to V} = C_{X \to Y}$ . In other words, linear transformation of the random variables will not change the srtength of causality between two variables $X$ and $Y$.

## RESULTS

### Type 1 Error of Statistics for Testing Causation

To examine the validity of statistics $T_C$ for testing the causal relationships between a common SNP and disease, we performed a series of simulation studies to compare their empirical levels with the nominal ones. We consier two scenarios: (1) no causation in the absence of association and (2) no causation in the presence of association. We selected the top 100 common SNPs (MAF between 0.19 and 0.49) from gene TEKT4P2 on chromosome 21 from 1,000 Genome Project. In scenario (1), a binary trait $Y$ is randomly generated and independent of indicator variables $X$ for genotypes of SNPs. In senario (2), we first randomly generated $X$ and $Y$, and then selected the associated pairs of data as our dataset $(X, Y)$.

We generated the data with 100,000 subjects by resampling from the 99-individual CEU population in 1,000 Genome Project. Number of permutations was 1,000, Number of replication of tests was 1,000. The sampled subjects from the generated population for type 1 error rate calculations were 500, 1,000, 2,000 and 5,000, respectively. We first consider scenatior 1. Table 2 summarized the average type I error rates of the test statistics for testing the causal relationsips between SNP and disease in the absence of association between SNP and disease over all 100 SNPs at the nominal levels $\alpha = 0.05$ and $\alpha = 0.01$ , respectively. To ensure no association in the data, we also presented Table 3 that summarized average type 1 error rates of the association test over 100 SNPs. These tables showed that the in the absence of association, type I error rates of the test statistics for testing the causal relationships between SNPs and disease were not appreciably different from the nominal levels. Next we consider scenartio 2. Table 4 presented the average type I error rates of the test statistics for testing the causal relationsips between SNP and disease in the presence of association between SNP and disease over all 100 SNPs at the nominal levels $\alpha = 0.05$ and $\alpha = 0.01$, respectively. Agan, these results demonstrated even in the presence of association the type I error rates of the test statistics for testing the causal relationships between SNPs and disease were not appreciably different from the nominal levels.

**Power Evaluation**

To evaluate the performance of the ANMs for assessing the causal relationships between SNP and disease, simulated data were used to estimate their power to detect a true causation. First, we invesitigate the power as a function of sample sizes with fixed causal measure parameter. The data were generated by the following cyclic model:

$$Y = f(X) + N_Y, N_Y \perp\!\!\!\perp X, \quad (15)$$

where $Y = \{0, 1\}$ *was* a binary trait and genearted by the model equation (15), $X = \{0, 1, 2\}$ was an indicator function for genotype of a SNP selected from 1,000 Genome Project, the minor allele frequency of the SNP was 0.1, $f$ was an integer function: $f(0) = 0, f(1) = 0, f(2) = 1$, $N_Y = \{0, 1\}$ was a noise distributed as a binormial with probability parameter $P$. We used the model equation (15) to generate the population of 100,000 individuals with $Y$ and $X$. A set of 500, 1,000, 2,000, 5,000, 10,000 and 20,000 individuals were sampled from the population. A total of 1,000 simulations were repeated for the power calculation. Three factors: the probability parameter $P$ in the bionomial distribution, significance level $\alpha$ and sample sizes affect the power of the ANMs for testing causation. We first fixed the parameter $P$ and significance level $\alpha$. Figure 2 plotted the power curves as a function of sample sizes where four scenarios: : (1) $P = 0.2, \alpha = 0.05$; (2) $P = 0.2, \alpha = 0.01$; (3) $P = 0.4, \alpha = 0.05$ and (4) $P = 0.4, \alpha = 0.01$ were considered. We observed from Fiure 2 that for $P = 0.2, \alpha = 0.01$, we could reach 81% power even when sample sizes were only 500 and for $P = 0.4, \alpha = 0.01$, we still could reach 80% power when sample sizes were 5,000.

We then fixed sample sizes $n$ and significance level $\alpha$. Figures 3 and 4 showed the power curves of the causation test as a function of the parameter $P$ with significance levels $\alpha = 0.05$ and $\alpha = 0.01$, respectively. We observed that when the parameter $P$ increased, the power of the causal tests decreased. Indeed, the parameter $P$ determined the value of the residual $N_Y$, which in turn, influenced the causality measure. When the parameter $P$ was small, the values of the response variable $Y$ were mainly determined by causal $X$. As the parameter $P$ increased, the impact of the noise $N_Y$ on $Y$ increased and hence the causality measure decreased, in turn, the power of the causal tests decreased. Finally, when $P = 0.5$, with the equal probability, the noise $N_Y$ produced values 1 and 0, $Y$ was mainly determined by noise $N_Y$, the ANMs had alomost no power to detect causation.

## Application to Real Data Example

### GWCS of Schizophenia

To further evaluate its performance, the ANMs for testing causation were applied to the CATIE-MGS-SWD schizophrenia (SCZ) study dataset with 8,421,111 common SNPs typed in 13,557 individuals.

In both GWAS and GWCS, the $\chi^2$ test was used for association analysis. A Manhattan plot of GWAS and GWCS was shown in Figure 5. For viewins clarity, in the Manhattan plot of GWAS and GWCS, we only showed P-values of causal analysis ( in green color) and association analysis (in black and grey colors) of all SNPs with P-values $< 10^{-5}$. We observed that associated SNPs were quite uniformally distributed across the genome, but the causal SNPs concentrated only on some genome regions. This may indicate that the Causal SNPs contained more information than the associated SNPs.

Due to computational time limitation of permutations, a P-value for declaring significant causation was $10^{-6}$. In total, 245 SNPs in 29 genes showed significant causations with SCZ. The results were summarized in Supplemental Table 1 where the P-values of both causation and association tests were listed. The selected top 15 causal SNPs were listed in Table 5. Among them, 62 causal SNPs can be confirmed from the literature and four of them were on the typical 108 schizophrenia-associated genetic loci (Schezoprenia working group, 2014; Sullivan et al. 2007; Fatemi et al. 2011; Lei et al. 2013; Costas et al. 2013; Athanasiu et al. 2013; Misztak et al. 2018; Ren et al. 2011; Suzuki et al. 2003; Cho et al. 2015; Ide and Lewis 2010). We also conducted GWAS for this dataset. A total of 5,917 SNPs are associated with SCZ at the significance level of $10^{-6}$ and only 58 showed causation.

These resuts showed several remarkable features. First, we can observe some SNPs that showed both significant causation and association. For example, four SNPs: rs1324544, rs2829725, rs9931378 and rs12057989 showed both strong causation and association (Table 5).

Second, the number of causal SNPs was much smaller than the number of associated SNPs. Third, highly significantly asscociated SNPs may show no significant causation. Fourth, the SNPs that showed strong causation signals may not demonstrate association. For example, SNP rs12739344 in gene AKT3 showed strong causation (P-value < $10^{-6}$), but did not reach threshold P-value for association (P-value for association is $8.95 \times 10^{-6}$) . It is well kown that the genetic variation in the gene AKT3 is a top risk signal in schizophrenia and network analysis identified that AKT3 contributes to four of the pathways involved in SCZ (Howell et al. 2017). SNP rs10986439 in gene GABBR2 showed significant causation (P-value < $10^{-6}$), but no association with SCZ (P-value is 0.000458). Genetic-imaging analysis showed that gene GABBR2 was in neuron development, synapse organization and axon pathways which could affect cognition in schizophrenia (Luo et al. 2018). Fifth, proportion of SNPs showed both causation and association was small (36.3% of causal SNPs showed association and only 0.98% of associated SNPs schowed causation).

**Disease Prediction**

Genomic predictors and risk estimates for a large number of diseases can be constructed from SNPs. The traditional methods for developing genomic risk scores (GRS) utilize small numbers of SNPs, typically those identified as genome-wide significant association (Abraham and Inouye 2015). To evaluate the predicitive ability of causal SNPs and associated SNPs, we selected the top 245 causal SNPs (all P-values < $10^{-6}$) and top 245 associated SNPs for SCZ risk prediction. Logistic regression and 10 fold cross validation were used to calculate prediction accuracy. Table 6 listed ten-fold cross-validated accuracy for prediction of SCZ. Table 6 showed that using the same number of SNPs, all the sets of SNPs selected by causal analysis had higher prediction accuracy than the set of SNPs selected by association analysis. Specifically, the prediction accuracy of 245 top causal SNPs was about 3% higher than that of the 245 top SNPs selected by association analysis. This may imply that the causal SNPs contain more biological information than associated SNPs.

**Impact of Linake Disequilibrium**

In this section, we investigate the impact of linkage disequilibrium (LD) on the causal analysis. It is well known that linkage disequilibrium has a large impact on the association analysis. The theoretical analysis of the impact of LD on the causal effect is gven in Supplementary E.

Next we use simulations to invesitigate the impact of LD on the causation analysis. Data for two markers: rs150012736 and rs376953511 were taken from 1000 Genome Project. In the 1000 Genome Project dataset , LD ($r^2$) between rs150012736 and rs376953511 was calculated as 0.5. Assume that SNP1 was a causal SNP. We did not make assumptions about whether or not SNP2 was a causal SNP. The trait values was generated by the discrete cyclic ANMs:

$$Y = f_Y(X) + N_Y, \qquad (16)$$

where $f_Y$ is a specified nonlinear integer function and $N_Y$ is a bionomial variable. We fitted the ANMs to the data $(Y, X^m)$ where $X^m$ represented the indicator variable for genotypes of SNP2. The results of causation and association tests were summarized in Tables S3 and S4, and Tables 7 and 8. Tables S3 and S4 showed that we can detect both association and causation between SNP1 and disease with a high power when sample sizes were larger than 2,000. Table 7 showed that type 1 error rates of test to detect causation between SNP2 and disease was not very high and decreased when sample sizes increased. In other words, we did not detect causation at SNP2. However, Table 8 showed that association test detected association of SNP2 with disease with high power.The simulation results showed that the impact of LD on the causal tests was much smaller than on the association tests.

To further evakuate the impact of LD on causation test, real data analysis was conducted. From the results of GWCS of SCZ, we selected SNP rs6578689 that had P-values < $10^{-6}$ and $2.82 \times 10^{-7}$ for causation and association tests, respectively. Then, we selected 20 neighboring SNPs of causal SNP rs6578689. We tested their causation and association with SCZ. Table 9 summarized the results of the causation and association tests. These results showed that even neighboring SNPs that had $r^2 > 0.44$ demonstrated no causation with SCZ, but strong associations with small P-values < $4.59 \times 10^{-9}$ with SCZ. These results of real data analysis demonstrated that LD had a small impact on causation analysis, but large impact on association tests.

**DISCUSSION**

Alternative to GWAS, the major goal of this paper is to propose a notion of GWCS and to address several important issues for GWCS. The standard approach to causal discovery is to use interventions or randomized experiments. Many genetic epidemiologists have always thought it impossible to detect causal SNPs using observational data. However, intervension or randomized experiments are unethical, time-consuming, expensive and infeasible in many cases. To address this critical barrier in GWCS, we focus on causal discovery methods developed for causal inference from observational data, not from interventional or randomized experiments and propose to use discrete ANMs as a major tool for GWCS. By large simulations and real data analysis we demonstrate the feasibility and limitations of the proposed GWCS as a new paradigm of genetic analysis.

Association is to measure dependent relationships and association analysis can be deone from observational data. Causal inference is inductive reasoning (Causal inference in AI, 2019). In other words, causal inference is reasonin from the observed part to the unobserved general. The goal of causal inference is to learn the response of taking an action and is usually carried out from interventions. However, as we pointed out before, it is infeasible to conduct intervention experiments in humans. Modern causal theory attempts to learn the outcome of an intervention from the observed data. Causation that can be inferred from observational data has been debated for more than a century. In this paper, we review great progresses that have been made in causal inferences over the past several decades, and define causation as the effect of taking action in some system from observational data in terms of interventions or counterfactuals (Lattimore and Ong 2018). We also review three emerging major approaches to bivariate causal discovery: "*do*" action, counterfactuals and ICM and showed that these three approaches can be unified. The ANMs that are widely used algorithms to implement ICM are explored for GWCS. In GWCS, we assume that there are no confoundings and selection bias. Methods for causation analysis with confounders will be presented elsewhere. Therefore, we lay down theoretic foundations for GWCS.

The original ANMs are used to distinguish cause-effect direction and do not provide P-value calculation for testing the causation of the SNP with disease. To overcome this limitation, we develop a test statistic and use permutations to calculate the P-value of statistics for testing the causation of the SNP with disease. This provides a practical approach to GWCS.

An essential problem for performing GWCS in practice is the type 1 error rates, power of the test statistics and feasibility of computations. We showed that type 1 error rates of the ANMs for testing the causation in both presence and absence of association were not significantly deviated from the nominal level. In other words, large simulation results demonstrated that the ANMs for causation analysis of genetic variants were valid. Power of the ANMs depends on the probability parameter $P$ in the bionomial distribution generating noise $N_Y$, sample sizes and significance levels. As we discussed in the text, probability parameter $P$ determines the strength of causation. We showed that even for significance lelvel $\alpha = 0.01$ and $P = 0.4$, when sample sizes were 5,000, the power of the ANMs was close to 80%. If the parameter $P \leq 0.15$, using 500 sample sizes, we could ensure that the ANMs can reach power greater than 90% under both $\alpha = 0.05$ and $\alpha = 0.01$. These results implied that the ANMs had high power to detect causation in many cases.

Distuinguishing causation from association is an age-old problem. The most classical causal inference theory focuses on inferring causal relationships among more than three variables. Due to lack of methods for bivariate causal discovery, very few GWCS and very few results of significant causal genetic variants from GWCS have been reported. In the past decade, the rapid development in modern causal analysis theory has provided several efficient methods for biovariate causal discovery including ANMs. To promote application of causal inference to genetic analysis, we applied the ANMs to GWCS of SCZ. From the GWCS of SCZ, we have several important observations.

First we observed that the number of causal SNPs (245 SNPs) was much less than the number of associated SNPs (5,917 SNPs). The cusal SNPs were mainly located in Chromosomes 1, 4, 5, 6, 7, 8, 20, 11, 12 and very few causal SNPs were located in other chromosomes. However, the associated SNPs were located across the genome. The results of GWCS of SCZ also challenged the "Omnigenic" model that assumed that "all genes affect every complex trait" (Greenwood 2018) and most association signals that tend to be spread across most of the genome influenced the phenotype variation (Boyle et al. 2017). The most identified association signals may have nothing to do with causing phenotype variation.

Second, the proportion of SNPs that showed both causation and association was small (36.3% of causal SNPs showed association and only 0.98% of associated SNPs schowed causation). This implied that the majority of causal SNPs could not be discovered by association analysis and most associated SNPs were not involved in the mechanisms of diseases. The results of GWCS of SCZ strongly suggested that association analysis will miss the majority of the causal SNPs and identifying and validating causal SNPs from the set of associated SNPs will be time consuming and not be efficient.

Third, full genomic information and genomic risk prediction has enabled new insights about the etiology and genetic architecture of complex disease. Although, we cannot directly validate the causality of the identified SNPs from GWCS, evaluating the difference in disease risk prediction accuracy between the set of causal SNPs and the set of associated SNPs allows assessing the biological relevance of the causal SNPs and associated SNPs. The prediction accuracy of 245 top causal SNPs was about 3% higher than that of 245 top associated SNPs. This may suggest that the causal SNPs contain more biological information than associated SNPs.

Fourth, both simulation and real data analysis showed that the LD had strong impact on association analysis, but surprisingly much less impact on the causal analysis. It is well known that LD is a confounding factor for association analysis and often creats spurious associations. Presence of LD across the genome will limit our abaility of using association analysis to discover mechanism of disease. Due to the limited impact of LD on causal analysis, we may expect that GWCS will provide an alternative to association analysis to discover causal genetic structure of complex diseases.

Although 62 of 245 discovered causal SNPs can be confirmed from the literature and four of them are on the typical 108 schizophrenia-associated genetic loci (Schizophrenia working group, 2014), the results were very preliminary. Functional studies of causal SNPs should be investigated in the future.

Causality is not only critical for us to understand disease mechanisms, but also particularly important for the development of efficient treatment. Much of the failure of previous efforts of drug development was attributable to the insufficient understanding of the disease mechanism.

The question whether we can infer causal relationships between genetic variants and disease from observational data has been debated for more than a century. Association and correlation analysis are the current paradigm of most genetic studies and have been used for more than a century. Our study demonstrated that large proportions of causal loci cannot be discovered by association analysis. Finding causal SNPs only via searching the set of associated SNPs may not be sufficient for unravelling mechanisms of complex diseases. Causal analysis as an alternative to association analysis for genetic studies has neven been systematically investigated. The main purpose of this paper is to stimulate discussion about causal analysis and association analysis, and both theoretical and practical research in genomic causal analysis. We hope that our results will greatly increase confidence in applying causal inference to genetic analysis, more and more intelligent methods for causal inference will be developed, and more and more valid GWCS of complex diseases will be investigated.

## DATA ACCESS

Software for implementing the proposed methods for GWCS can be downloaded from https://sph.uth.edu/research/centers/hgc/xiong/software.htm and Github ( https://github.com/jiaorong007?tab=repositories ) .

Table 1. Contingency table for testing independence.

| | Genotype $dd$ | Genotype $dD$ | Genotype $DD$ | |
|---|---|---|---|---|
| $N_Y = 0$ | $a_{11}$ | $a_{12}$ | $a_{13}$ | $n_0$ |
| $N_Y = 1$ | $b_{1\,1}$ | $b_{12}$ | $b_{13}$ | $n_1$ |
| | $a_{1\,1} + b_{1\,1}$ | $a_{12} + b_{12}$ | $a_{13} + b_{13}$ | $n = n_0 + n_1$ |

Table 2. Average type 1 error rates of the statistics for testing causal relationships between SNP and disease.

| | Sample Size | | | |
|---|---|---|---|---|
| Nominal Level | 500 | 1,000 | 2,000 | 5,000 |
| 0.05 | 0.044 | 0.046 | 0.048 | 0.051 |
| 0.01 | 0.005 | 0.006 | 0.007 | 0.009 |

Table 3. Type 1 error rates for association test.

| Nominal Level | 500 | 1,000 | 2,000 | 5,000 |
|---|---|---|---|---|
| 0.05 | 0.05 | 0.05 | 0.049 | 0.049 |
| 0.01 | 0.01 | 0.01 | 0.01 | 0.01 |

Table 4. Average type 1 error rates of the statistics for testing causal relationships between SNP and disease in the presence of association.

| Nominal Level | 500 | 1,000 | 2,000 | 5,000 |
|---|---|---|---|---|
| 0.05 | 0.042 | 0.046 | 0.047 | 0.046 |
| 0.01 | 0.005 | 0.007 | 0.007 | 0.008 |

Table 5. P-values of top 15 SNPs that had significant causal relationships with schizophrenia.

| | | | | | P-values | |
|---|---|---|---|---|---|---|
| RS Number | Chr | Position | Gene | Related Disease | Causation | Association |
| rs1324544 | 6 | 9181479 | | | <E-06 | 3.14E-12 |
| rs2829725 | 21 | 26764027 | | | <E-06 | 4.53E-11 |
| rs9931378 | 16 | 5783022 | | | <E-06 | 1.23E-09 |
| rs12057989 | 1 | 144617251 | | | <E-06 | 1.28E-08 |
| rs7110863 | 11 | 112843138 | NCAM1 | Schizophrenia | <E-06 | 4.34E-08 |
| rs1420643 | 7 | 35874928 | SEPT7 | Schizophrenia | <E-06 | 2.02E-07 |
| rs1534440 | 6 | 145017328 | UTRN | Schizophrenia | <E-06 | 2.36E-07 |
| rs228768 | 17 | 42191893 | HDAC5 | Mental Depression | <E-06 | 3.76E-06 |
| rs1940713 | 11 | 112906285 | NCAM1 | Schizophrenia | <E-06 | 4.57E-06 |
| rs1940714 | 11 | 112906391 | NCAM1 | Schizophrenia | <E-06 | 4.57E-06 |
| rs12739344 | 1 | 243791312 | AKT3 | Schizophrenia | <E-06 | 8.95E-06 |
| rs876983 | 8 | 18407858 | PSD3 | Schizophrenia | <E-06 | 1.42E-05 |
| rs10075211 | 5 | 147839537 | HTR4 | Schizophrenia | <E-06 | 2.24E-05 |
| rs725515 | 16 | 82854696 | CDH13 | Mental Depression | <E-06 | 3.80E-05 |
| rs10986439 | 9 | 101262400 | GABBR2 | Major Depressive Disorder | <E-06 | 0.000457917 |

Table 6. Ten-fold cross-validated accuracy and AUC for SCZ risk prediction of using top 15 causal SNPs and association SNPs.

| Number of SNPs | 7 | 8 | 9 | 10 | 11 | 12 | 13 | 14 | 15 | 245 |
|---|---|---|---|---|---|---|---|---|---|---|
| Accuracy of Causal SNPs | 0.5511 | 0.5542 | 0.5542 | 0.5542 | 0.5542 | 0.5540 | 0.5534 | 0.5531 | 0.5521 | 0.5737 |
| AUC of Causal SNPs | 0.5320 | 0.5340 | 0.5343 | 0.5345 | 0.5344 | 0.5342 | 0.5336 | 0.5333 | 0.5324 | 0.5491 |
| Accuracy of Associated SNPs | 0.5470 | 0.5457 | 0.5434 | 0.5423 | 0.5415 | 0.5410 | 0.5404 | 0.5401 | 0.5395 | 0.5430 |
| AUC of Associated SNPs | 0.5204 | 0.5200 | 0.5191 | 0.5189 | 0.5178 | 0.5173 | 0.5168 | 0.5163 | 0.5158 | 0.5249 |

Table 7. Type I error rates of causal test between SNP2 and disease.

| Significance Level | 500 | 1,000 | 2,000 | 5,000 |
|---|---|---|---|---|
| 0.05 | 0.183 | 0.159 | 0.142 | 0.104 |
| 0.01 | 0.105 | 0.118 | 0.105 | 0.093 |

Table 8. Power of test for association between SNP2 and disease.

| Significance Level | 500 | 1,000 | 2,000 | 5,000 |
|---|---|---|---|---|
| 0.05 | 0.918 | 0.979 | 0.992 | 0.994 |
| 0.01 | 0.860 | 0.957 | 0.990 | 0.992 |

Table 9. P-values for causation and association tests of 20 neighboring SNPs of causal SNP rs6578689.

| SNPs | Chr | P-values | | Neighbor SNPs | Position | $r^2$ | P-values | |
|---|---|---|---|---|---|---|---|---|
| | | Causation | Association | | | | Causation | Association |
| rs6578689 | 11 | <E-06 | 2.82E-07 | rs10742794 | 5826464 | 0.7196 | 0.03 | 9.95E-10 |
| rs6578689 | 11 | | | rs11039135 | 5836787 | 0.63226 | 0.39 | 2.65E-09 |
| rs6578689 | 11 | | | rs7115498 | 5831847 | 0.53094 | 0.94 | 9.31E-10 |
| rs6578689 | 11 | | | rs10838661 | 5830617 | 0.53093 | 0.96 | 1.03E-09 |
| rs6578689 | 11 | | | rs35898746 | 5830823 | 0.53093 | 0.96 | 1.03E-09 |
| rs6578689 | 11 | | | rs11039085 | 5823651 | 0.53034 | 0.93 | 6.03E-10 |
| rs6578689 | 11 | | | rs10742791 | 5819152 | 0.5272 | 0.94 | 4.03E-10 |
| rs6578689 | 11 | | | rs12226188 | 5837141 | 0.52658 | 0.9 | 6.80E-10 |
| rs6578689 | 11 | | | rs10838674 | 5836857 | 0.52634 | 0.93 | 9.01E-10 |
| rs6578689 | 11 | | | rs35271555 | 5833707 | 0.5233 | 0.9 | 7.99E-10 |
| rs6578689 | 11 | | | rs6578687 | 5813985 | 0.52136 | 0.95 | 5.00E-10 |
| rs6578689 | 11 | | | rs7114690 | 5814376 | 0.51743 | 0.88 | 3.83E-10 |
| rs6578689 | 11 | | | rs80316576 | 5827945 | 0.44329 | 0.37 | 3.01E-09 |
| rs6578689 | 11 | | | rs73390385 | 5809052 | 0.44286 | 0.42 | 3.96E-09 |
| rs6578689 | 11 | | | rs73392251 | 5821745 | 0.44191 | 0.44 | 4.59E-09 |
| rs6578689 | 11 | | | rs73392254 | 5822797 | 0.44191 | 0.54 | 4.59E-09 |
| rs6578689 | 11 | | | rs73390383 | 5808495 | 0.44143 | 0.48 | 3.90E-09 |
| rs6578689 | 11 | | | rs73392222 | 5817732 | 0.44136 | 0.47 | 2.80E-09 |
| rs6578689 | 11 | | | rs73392226 | 5817797 | 0.44136 | 0.46 | 2.80E-09 |
| rs6578689 | 11 | | | rs77107630 | 5818487 | 0.44136 | 0.47 | 2.80E-09 |

## Figure Legends

**Figure 1.** Several possible causal relationships between two observed variables $X$ and $Y$: (a) association; (b) $X$ causes $Y$; (c) $Y$ causes $X$; (d) temperature change causes thermometer change.

**Figure 2.** Power curves of the ANMs for testing causation as a function of sample sizes where power of the tests was calculated under four scenarios: (1) $P = 0.2, \alpha = 0.05$; (2) $P = 0.2, \alpha = 0.01$; (3) $P = 0.4, \alpha = 0.05$ and (4) $= 0.4, \alpha = 0.01$.

**Figure 3.** Power convers of the ANMs for testing causation as a function of the parameter $P$ in bionomial distribution where four sample sizes 500, 1,000, 5,000 and 10,000 were considered, assuming $\alpha = 0.05$.

**Figure 4.** Power curves of the ANMs to test causation as a function of the parameter $P$ in bionomial distribution. Four sample sizes 500, 1,000, 5,000 and 10,000 were considered, assuming $\alpha = 0.01$.

**Figure 5.** A Manhattan plot of GWAS and GWCS.

Figure 1.

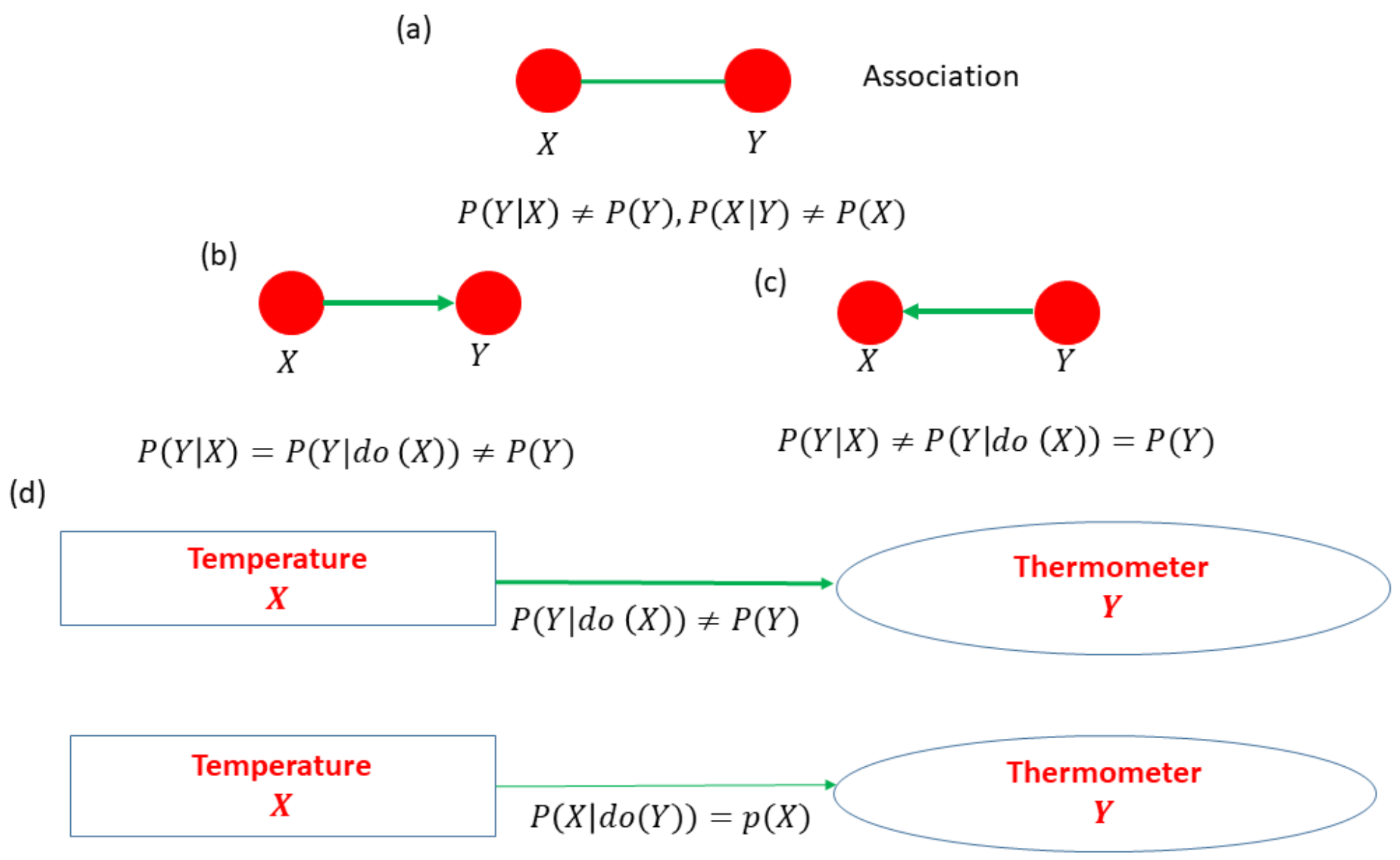

Figure 2.

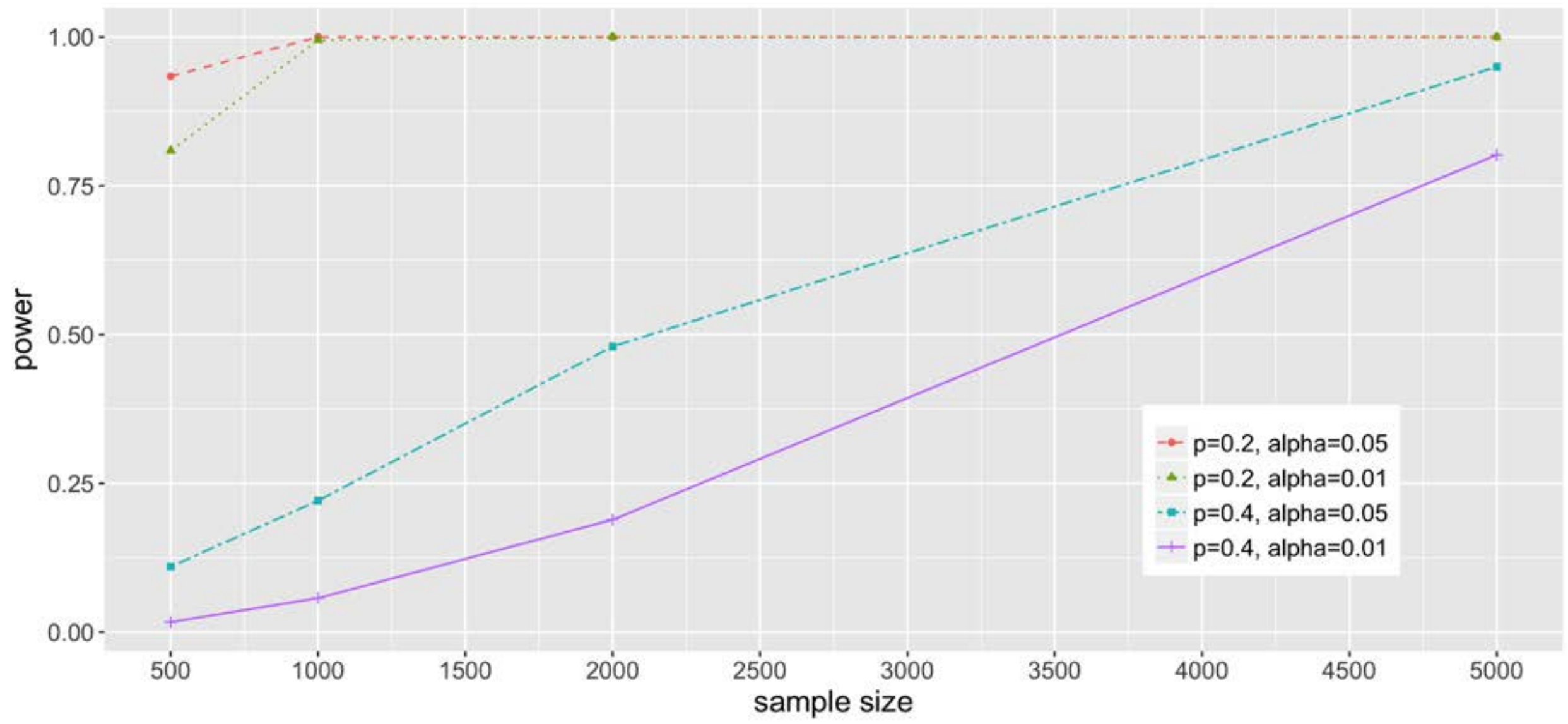

Figure 3.

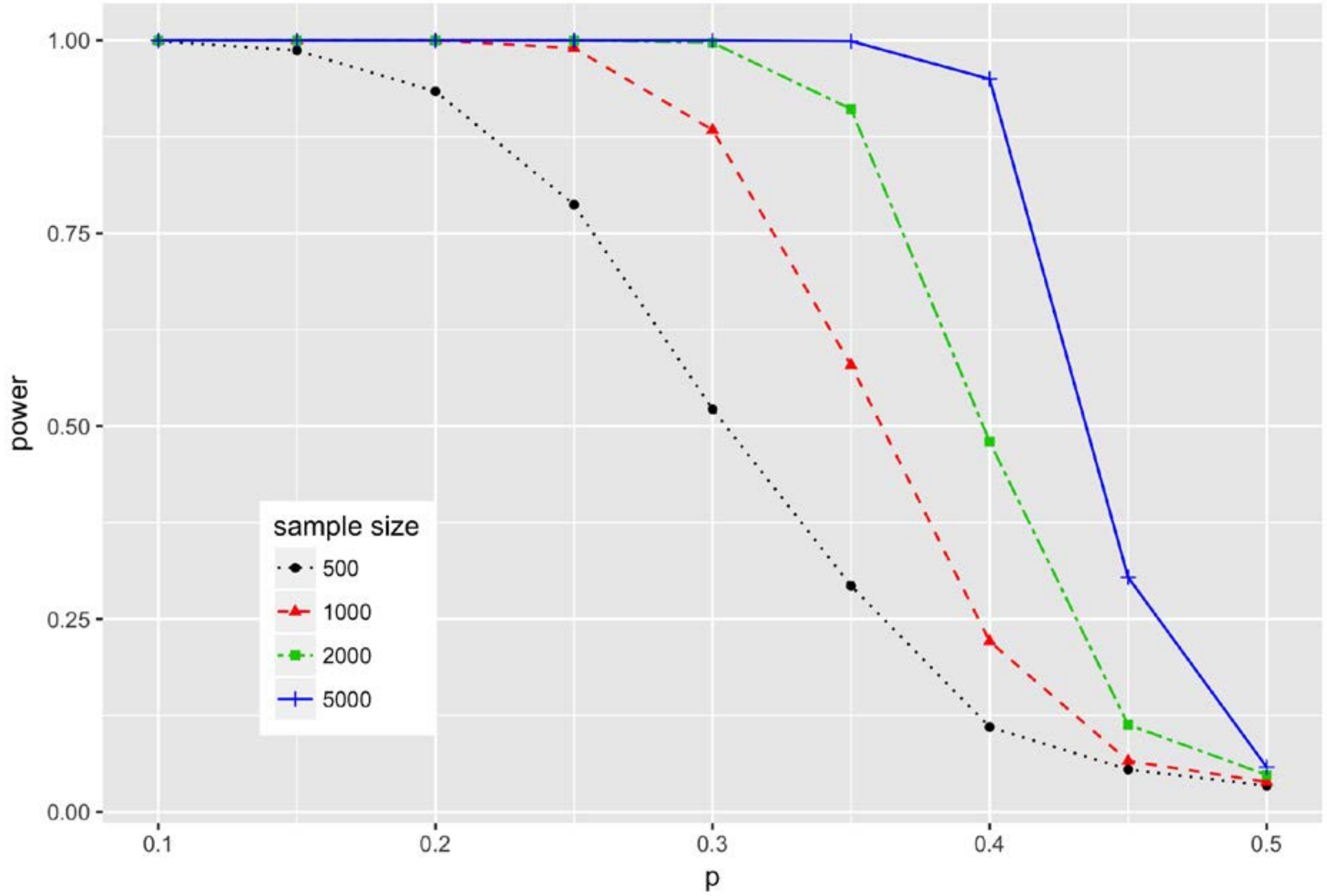

Figure 4.

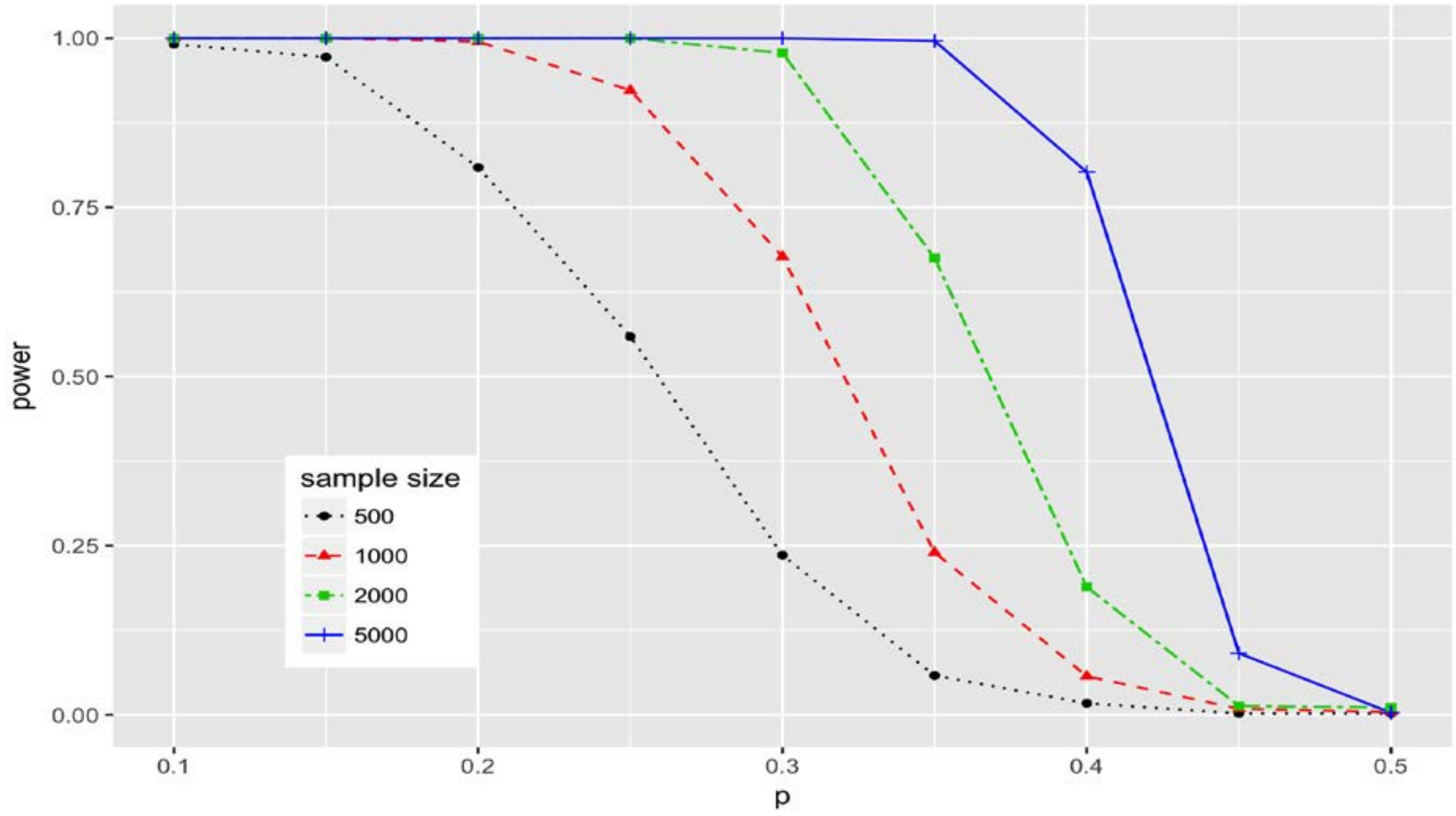

Figure 5.

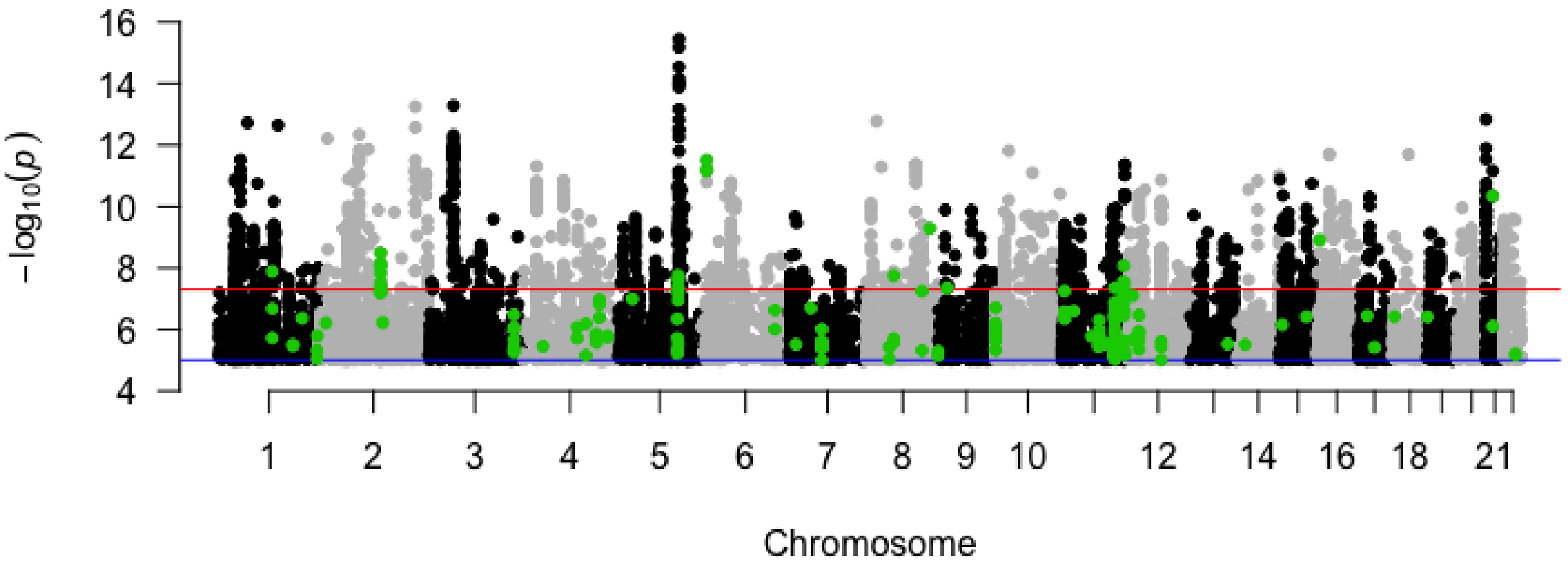

**Supplementary A**

**Counterfactuals for causal inference**

This introduction focuses on how to use counterfactuals to investigate the causal effect. The average causal effect (ACE) (or treatment effect) is defined as

$$\text{ACE}= E[Y_i^1] - E[Y_i^0], \tag{A1}$$

where $E[.]$ is taken over entire population (Elwert 2013) .

Since for each individual we can only observe one of $Y_i^1$ and $Y_i^0$, we cannot estimate ACE. Standard statistics to estimate ACE is given by

$$S = E[Y_i^1 | T_i = 1] - E[Y_i^0 | T_i = 0]. \tag{A2}$$

where $E[.]$ is taken over the treatment group and control group, respectively, not over the entire population. If potential outcome is binary, then equations (A1) and (A2) can be rewritten as

$$ACE = P(Y_i^1 = 1) - P(Y_i^0 = 1) \tag{A3}$$

and

$$S = P(Y_i^1 = 1 | T_i = 1) - P(Y_i^0 = 1 | T_i = 0) . \tag{A4}$$

Since quantity $S$ depends on the treatment assignment, it measures the association between the potential outcome with the treatment asigment. Therefore, in general, the ACE is not equal to $S$. The sufficient conditions to make them equal are

Condition I:

$$E[Y_i^1] = E[Y_i^1 | T_i = 1] = E[Y_i^1 | T_i = 0] \quad \text{(A5)}$$

or

$$P(Y_i^1 = 1) = P(Y_i^1 | T_i = 1) = P(Y_i^1 | T_i = 0), \quad \text{(A6)}$$

which implies that the mean potential outcome (or the probability distribution) under treatment for those in the treatment group equals the mean potential outcome (or the probability distribution) under treatment for those in the control group.

Condition II:

$$E[Y_i^0] = E[Y_i^0 | T_i = 1] = E[Y_i^0 | T_i = 0] \quad \text{(A7)}$$

or

$$P(Y_i^0 = 1) = P(Y_i^0 = 1 | T = 1) = P(Y_i^0 = 1 | T = 0) , \quad \text{(A8)}$$

which implies that the mean potential outcome (or the probability distribution) under control for those in the treatment group equals the mean potential outcome (or the probability distribution) under control for those in the control group.

Under conditions I and II we obtain

$$S = E[Y_i^1 | T = 1] - E[Y_i^0 | T = 0] = E[Y_i^1] - E[Y_i^0] = ACE , \quad \text{(A9)}$$

or

$$S = P(Y_i^1 = 1 | T = 1) - P(Y_i^0 = 1 | T = 0) = P(Y_i^1 = 1) - P(Y_i^0 = 1) = ACE. \quad \text{(A10)}$$

In other words, conditions I and II ensure that association measure $S$ is equal to the average causal effect ACE. Conditions I and II imply and assume that the average potential outcomes of people in the treatment group are equal to that of the outcomes of the people in the control group. In other words, no difference between association measurement and causal measurement can be achieved by randomized treatment assignment.

Random experiments are often expensive, unethical and infeasible. Observational data should be used for causal inference. The assumption to ensure that $S$ is an unbiased and consistent estimator of the ACE is the following ignorability:

$$(Y_i^1, Y_i^0) \perp\!\!\!\perp T, \qquad \text{(A11)}$$

i.e., the potential outcomes must be jointly independent of treatment assignment.

In the observational studies, the ignorability assumption, in general, is difficult to be satisfied. Therefore, we make further assumptions to extend the ignorability to conditional ignorability:

$$(Y_i^1, Y_i^0) \perp\!\!\!\perp T|X , \qquad \text{(A12)}$$

where $X$ is a set of variables.

Conditional ignorability in equation (A12) assumes that the potential outcomes, $Y_i^1$ and $Y_i^0$ are jointly independent of treatment assignment conditional on groups defined by the value of $X$.

**Supplementary B**

**Unification of the SEMs, ICM, counterfactuals and do-calculus methods for causal inference**

In this supplementary, we briefly show that the SEMs, ICM, counterfactuals and do-calculus methods for causal inference with two random variables can be unified.

Suppose that both $X$ and $Y$ are binary variables. Taking action $X = 1$ implies that the equation $X = f_x(\varepsilon_x)$ should be replaced by $X = 1$. Setting $X = 1$ will not affect the function $f_Y$ and distribution of noise $\varepsilon_Y$. The interventional distribution is given by

$$P\left(Y = 1 \middle| do(X = 1)\right) = \sum_{\varepsilon_Y} P(\varepsilon_Y) I\{f_Y(1, \varepsilon_Y) = y\}. \quad \text{(B1)}$$

Note that the conditional observational distribution of $Y$ given $X$ is

$$P(Y = 1|X = 1) = \sum_{\varepsilon_x} \sum_{\varepsilon_Y} P(\varepsilon_x|X = 1) P\left(\varepsilon_y \middle| \varepsilon_x\right) I\{f_y\left(1, \varepsilon_y\right) = y\} . \quad \text{(B2)}$$

By the assumption $\varepsilon_x \perp\!\!\!\perp \varepsilon_Y$, we have

$$P\left(\varepsilon_y \middle| \varepsilon_x\right) = P(\varepsilon_y). \quad \text{(B3)}$$

Substituting equation (B3) into equation (B2) yields

$$P(Y = 1|X = 1) = \sum_{\varepsilon_y} P(\varepsilon_y) I\{f_y\left(1, \varepsilon_y\right) = y\}. \quad \text{(B4)}$$

Combining equations (B1) and (B4), we obtain

$$P\left(Y = 1 \middle| do(X = 1)\right) = P(Y = 1|X = 1), \quad \text{(B5)}$$

which shows that interventional distribution $P\left(Y = 1 \middle| do(X = 1)\right)$ is equal to the observational distribution $P(Y = 1|X = 1)$.

Similarly, we can prove

$$P\big(Y = 1\big|do(X = 0)\big) = P(Y = 1|X = 0). \tag{B6}$$

The causal effect of $X$ on the variable $Y$ under the structural equation model is defined as

$$ACES = P\big(Y = 1\big|do(X = 1)\big) - P\big(Y = 1\big|do(X = 0)\big). \tag{B7}$$

Equations (B5) and (B6) show that under the structural equation model (4) the interventional distribution is equal to the observational distribution.

Under the ignorability assumption $(Y_i^1, Y_i^0) \perp\!\!\!\perp X_i$, we obtain

$$P(Y_i^1 = 1) = P(Y_i^1 = 1\big|X_i = 1) = P(Y = 1|X = 1), \tag{B8}$$

$$P(Y_i^0 = 1) = P(Y_i^0\big|X_i = 1) = P(Y = 1|X = 0). \tag{B9}$$

Combining equations (B5), (B6), (B7) and (B8), we obtain

$$P\big(Y = 1\big|do(X = 1)\big) = P(Y_i^1 = 1), \tag{B10}$$

$$P\big(Y = 1\big|do(X = 0)\big) = P(Y_i^0 = 1). \tag{B11}$$

It follows from equations (A3), (B7), (B10) and (B11) that

$$ACES = P\big(Y = 1\big|do(X = 1)\big) - P\big(Y = 1\big|do(X = 0)\big) = P(Y_i^1 = 1) - P(Y_i^0 = 1) = ACE.$$

(B12))

Equation (B12) shows that the causal effect under the SEMs is equal to the average causal effect under the counterfactual model with ignorability assumption.

Next we discuss the equivalence between ICM and SEMs. Take the SEMs (4) as transformations:

$$X = f_x(\varepsilon_x) \,,$$

$$Y = f_y(X, \varepsilon_Y) \,.$$

We need to show that ICM of $X \rightarrow Y$ implies $\varepsilon_x \perp\!\!\!\perp \varepsilon_Y$.

The Jacobian matrix of the above transformation is

$$J = \frac{\partial f_x}{\partial \varepsilon_x}\frac{\partial f_y}{\partial \varepsilon_y} \,.$$

Then, by transformation theorem (Ross 1985), we obtain

$$P(X,Y) = \frac{P(\varepsilon_x, \varepsilon_y)}{|\frac{\partial f_x \partial f_y}{\partial \varepsilon_x \partial \varepsilon_y}|} \,, \tag{B13}$$

$$P(X) = \frac{P(\varepsilon_x)}{|\frac{\partial f_x}{\partial \varepsilon_x}|} \,, \tag{B14}$$

which implies that

$$P(Y|X) = \frac{\frac{P(\varepsilon_x, \varepsilon_y)}{|\frac{\partial f_x \partial f_y}{\partial \varepsilon_x \partial \varepsilon_y}|}}{\frac{P(\varepsilon_x)}{|\frac{\partial f_x}{\partial \varepsilon_x}|}} = \frac{P(\varepsilon_x, \varepsilon_y)}{P(\varepsilon_x)|\frac{\partial f_y}{\partial \varepsilon_y}|}. \tag{B15}$$

ICM states that distributions $P(X)$ and $P(Y|X)$ are independent, or $P(Y|X)$ contains no information of $P(X)$. Therefore, $P(\varepsilon_x, \varepsilon_y)$ must be equal to $P(\varepsilon_x)P(\varepsilon_y)$, i.e., $\varepsilon_x \perp\!\!\!\perp \varepsilon_Y$. Otherwise, from equation (B15), we know that $P(Y|X)$ involves both $\varepsilon_x$ and $\varepsilon_y$, which implies the distribution $P(Y|X)$ will contain information of $P(X)$ , and ICM will not hold. This shows that ICM of $X \rightarrow Y$ implies the structural equation model (4).

**Supplementary C**

**Identifiability of the direction of discrete ANMs**

Conditions for identifiability were summarized in Theorem 4 in the paper (Peters et l. 2011) . They assumed that $Y = f_y(X) + N_y,\ X \perp\!\!\!\perp N_y, P(X = k) \neq 0, P(N_Y = l) \neq 0, \forall k, l$ and considered three cases:

(1) Both $f_Y$ and $f_X$ are bijective. If the ANM $X \rightarrow Y$ is reversible, then $X$ and $Y$ are uniformly distributed;

(2) $f_X$ is bijective. Suppose that $f_X(Y_1) = f_X(Y_2)$. If the ANM $X \rightarrow Y$ is reversible, then $\frac{P(N_Y=Y_1-f_Y(X))}{P(N_Y=Y_2-f_Y(X))} = \frac{P(Y=Y_1)}{P(Y=Y_2)}, \forall X,$ in many cases, $P(Y = Y_1) = P(Y = Y_2)$.

(3) $f_Y$ is bijective. Suppose that $f_Y(X_1) = f_Y(X_2)$. If the ANM $X \rightarrow Y$ is reversible, then $\frac{P(X=X_1)}{P(X=X_2)} = \frac{P(N_X=X_1-f_X(Y))}{P(N_X=X_2-f_X(Y))}, \forall Y$, in many cases, $P(X = X_1) = P(X = X_2)$ .

In our cases, $X$ is an indicator variable for genotypes and $Y$ is a binary variable for disease status. Therefore, in general, in any of the three cases, reversible is impossible and hence the direction of the ANMs is identifiable.

**Supplementary D**

**Distance covariance and correlation between two random vectors**

The distance covariance $dCov^2(X, Y)$ between two random vectors $X$ and $Y$ with finite first moments is defined as

$$dCov^2(X,Y) = ||f_{X,Y}(t,s) - f_X(t)f_Y(s)||^2 = \frac{1}{c_p c_q}\int_{R^{p+q}} \frac{|f_{X,Y(t,s)} - f_X(t)f_Y(s)|^2}{|t|_p^{1+p}|s|_q^{1+q}} dtds \quad \text{(D1)}$$

where $c_p = \frac{\pi^{\frac{1+p}{2}}}{\Gamma(\frac{1+p}{2})}$ and $c_q = \frac{\pi^{\frac{1+q}{2}}}{\Gamma(\frac{1+q}{2})}$.

Similarly, distance variance $dVar^2(X)$ is defined as

$$dVar^2(X) = \frac{1}{c_p^2}\int_{R^{2p}} \frac{|f_{X,X(t,s)} - f_X(t)f_X(s)|^2}{\left|t\right|_p^{1+p}\left|s\right|_p^{1+p}} dtds\ . \quad \text{(D2)}$$

The square of distance correlation $R^2(X,Y)$is defined as

$$R^2(X,Y) = \begin{cases} \frac{dCov^2(X,Y)}{\sqrt{dVar^2(X)dVar^2(Y)}} & dVar^2(X)dVar^2(Y) > 0 \\ 0 & dVar^2(X)dVar^2(Y) = 0 \end{cases}. \quad \text{(D3)}$$

The distance covariance and correlation can be easily estimated as follows (Sze´kely et al. 2007). Assume that pairs of $(X_k, Y_K), k = 1,...,n$ are sampled. Calculate the Euclidean distances:

$a_{kl} = \| X_K - X_l \|_p$, $b_{kl} = \| Y_k - Y_l \|_q, k = 1,...,n, l = 1,...,n$ .

Define

$$\bar{a}_{k.} = \frac{1}{n}\sum_{l=1}^{n} a_{kl}\ ,\ \bar{a}_{.l} = \frac{1}{n}\sum_{k=1}^{n} a_{kl}\ ,\ \bar{a}_{..} = \frac{1}{n^2}\sum_{k=1}^{n}\sum_{l=1}^{n} a_{kl}\ ,$$

$$\bar{b}_{k.} = \frac{1}{n}\sum_{l=1}^{n} b_{kl}\ ,\ \bar{b}_{.l} = \frac{1}{n}\sum_{k=1}^{n} b_{kl} \text{ and } \bar{b}_{..} = \frac{1}{n^2}\sum_{k=1}^{n}\sum_{l=1}^{n} b_{kl}\ .$$

Define two matrices:

$$A = (A_{kl})_{n\times n} \text{ and } B = (B_{kl})_{n\times n},$$

where

$$A_{kl} = a_{kl} - \bar{a}_{k.} - \bar{a}_{.l} + \bar{a}_{..},$$

$$B_{kl} = b_{kl} - \bar{b}_{k.} - \bar{b}_{.l} + \bar{b}_{..},\ k,l = 1,\ldots,n.$$

Finally, the sampling distance covariance $V_n(X,Y)$, variance $V_n(X)$ and correlation $R_n(X,Y)$ are defined as

$$V_n^2(X,Y) = \frac{1}{n^2}\sum_{k=1}^{n}\sum_{l=1}^{n} A_{kl}B_{kl}, \qquad \text{(D4)}$$

$$V_n^2(X) = V_n^2(X,X) = \frac{1}{n^2}\sum_{k=1}^{n}\sum_{l=1}^{n} A_{kl}^2,\ V_n^2(Y) = \sum_{k=1}^{n}\sum_{l=1}^{n} B_{kl}^2,$$

$$R_n^2(X,Y) = \begin{cases} \dfrac{V_n^2(X,Y)}{\sqrt{V_n^2(X)V_n^2(Y)}}, & V_n^2(X)V_n^2(Y) > 0 \\ 0 & V_n^2(X)V_n^2(Y) = 0, \end{cases} \qquad \text{(D5)}$$

respectively.

## Supplementary E

## Theoretical Analysis of the Impact of LD on the Causal Effects

For the convenience of presentation, we first consider the true linear model for a quantitative trait (Xiong 2018):

$$Y = \mu + X\alpha + N_Y, X \perp\!\!\!\perp N_Y, \qquad \text{(E1)}$$

where $X$ is an inicater variable for the genotype at the true causal locus and distribution of $N_Y$ is not normal.

Suppose that $X^m$ is an indicator variable for the genotype at a marker locus with marker allele frequencies $P_M$ and $P_m$ and LD measure $D_m$ between the marker and true causal loci. Then, we have the following linear regression model for the marker locus:

$$Y = \mu + X^m\alpha_m + N_Y^m. \qquad \text{(E2)}$$

Then, we can show (Xiong 2018) that

$$\alpha_m \xrightarrow{a.s} \frac{D_m}{P_M P_m} \alpha. \tag{E3}$$

Equation (E3) implies that in the presence of LD, the marker locus still shows some association with genetic additive effect $\frac{D_m}{P_M P_m} \alpha$ approximately.

Now we investigate the impact of LD on causal inference. Substituting equation (E1) into equation (E2), we obtain

$$N_Y^m = N_Y + X\alpha - X^m \alpha_m \, . \tag{E4}$$

Define

$$\Delta = X\alpha - X^m \alpha_m \approx (X - \frac{D_m}{P_M P_m} X^m)\alpha. \tag{E5}$$

When $\Delta \neq 0$, distance covariance $dCov^2(X^m, N_Y^m)$ is equal to

$$\begin{aligned} 0 \le dCov^2(X^m, N_Y^m) &= dCov^2(X + X^m - X, N_Y + \Delta) \\ &\le dCov^2(X, N_Y) + dCov^2(X^m - X, \Delta) \\ &= dCov^2(X^m - X, \Delta), \end{aligned} \tag{E6}$$

where $X$ and $N_Y$ are independent by the ICM.

$X^m \to Y$ must imply that $\Delta = 0$ (Sze´kely and Rizzo, 2009) or

$$X = \frac{D_m}{P_M P_m} X^m. \tag{E7}$$

Equation (E7) indicates that $X^m \to X$. However, in general, SNPs do not have causal relationships. Therefore, $dCov^2(X^m, N_Y^m) \neq 0$ and $X^m, N_Y^m$ are not independent, which implies that $X^m$ does not cause $Y$.

Now we calculate the causal measure. Let $C_{X \to Y} = 1 - R(X, N_Y)$ be the causal measure of the causal SNP $X$. Then, the causal measure of the marker $X^m$ is given by

$$C_{X^m \to Y} = C_{X \to Y} - R(X^m - X, X - \frac{D_m}{P_M P_m} X^m). \tag{E8}$$

$1 \geq R(X^m - X, X - \frac{D_m}{P_M P_m} X^m) \geq 0$ implies

$$C_{X \to Y} \geq C_{X^m \to Y} \geq 0. \quad \text{(E9)}$$

Causation measure $C_{X^m \to Y}$ depends on the distance correlation between $X^m - X$ and $X - \frac{D_m}{P_M P_m} X^m$.

For qualitative trait, we can use a logistic integer function as a nonlinear function. After some algebraic operations, we have the model:

$$Y = \frac{e^{X\alpha}}{1+e^{X\alpha}} + N_Y \quad \text{(E10)}$$

or

$$Y = f(X\alpha) + N_Y , \quad \text{(E11)}$$

where $f(X\alpha)$ is a nonlinear function.

Equation (E11) can be approximated by

$$Y = f(0) + f'(0)X\alpha + N_Y . \quad \text{(E12)}$$

Thus, the model (E11) is reduced to model equation (E1). Using the same arguments from the model equation (E1), we can define the causality measure for marker $X^m$:

$$C_{X^m \to Y} = C_{X \to Y} - R(X^m - X, X - \frac{f'(0) D_m}{P_M P_m} X^m). \quad \text{(E13)}$$

For the discrete ANMs, we cannot find $f'(0)$, the causal measure for the marker may simply be written as

$$C_{X^m \to Y} = C_{X \to Y} - R(X^m - X, X - \frac{\gamma D_m}{P_M P_m} X^m), \quad \text{(E14)}$$

where $\gamma$ is a appropriate constant.

Table S1. P-values of causation and assication of 214 SNPs showing significant causal relationships with schizophrenia.

| SNPs | Chr | Gene | Related Disease | P-values | |
|---|---|---|---|---|---|
| | | | | Causation | Association |
| rs1324544 | 6 | | | <E-06 | 3.14E-12 |
| rs4715076 | 6 | | | <E-06 | 6.90E-12 |
| rs2829725 | 21 | | | <E-06 | 4.53E-11 |
| rs9931378 | 16 | | | <E-06 | 1.23E-09 |
| rs12057989 | 1 | | | <E-06 | 1.28E-08 |
| rs11784724 | 8 | | | <E-06 | 1.76E-08 |
| rs13167565 | 5 | ADAMTS19 | Premature Menopause | <E-06 | 2.33E-08 |
| rs1820353 | 11 | | | <E-06 | 2.83E-08 |
| rs1433979 | 11 | | | <E-06 | 3.09E-08 |
| rs11746536 | 5 | KIAA1024L | | <E-06 | 4.23E-08 |
| rs10791112 | 11 | | | <E-06 | 4.29E-08 |
| rs7110863 | 11 | NCAM1 | Schizophrenia (S2) | <E-06 | 4.34E-08 |
| rs990323 | 5 | | | <E-06 | 4.86E-08 |
| rs6578683 | 11 | | | <E-06 | 5.60E-08 |
| rs28817943 | 8 | | | <E-06 | 5.68E-08 |
| rs1118137 | 11 | | | <E-06 | 6.52E-08 |
| rs1465402 | 5 | ADAMTS19 | Premature Menopause | <E-06 | 6.64E-08 |
| rs929683 | 12 | | | <E-06 | 7.96E-08 |
| rs7942900 | 11 | | | <E-06 | 8.14E-08 |
| rs1347283 | 11 | | | <E-06 | 1.03E-07 |
| rs6822457 | 4 | | | <E-06 | 1.05E-07 |
| rs7118907 | 11 | NCAM1 | Schizophrenia (S2) | <E-06 | 1.32E-07 |
| rs961491 | 11 | | | <E-06 | 1.35E-07 |
| rs2212450 | 11 | | | <E-06 | 1.38E-07 |
| rs1481216 | 4 | | | <E-06 | 1.50E-07 |
| rs11251290 | 10 | | | <E-06 | 1.95E-07 |
| rs4937872 | 11 | | | <E-06 | 1.95E-07 |
| rs1420643 | 7 | SEPT7 | Schizophrenia (S2) | <E-06 | 2.02E-07 |
| rs9442081 | 1 | | | <E-06 | 2.10E-07 |
| rs1534440 | 6 | UTRN | Schizophrenia (S2) | <E-06 | 2.36E-07 |
| rs7122268 | 11 | | | <E-06 | 2.45E-07 |
| rs2090636 | 11 | | | <E-06 | 2.51E-07 |
| rs2324316 | 11 | | | <E-06 | 2.55E-07 |
| rs10892831 | 11 | | | <E-06 | 2.74E-07 |
| rs2186710 | 11 | | | <E-06 | 2.75E-07 |
| rs6578689 | 11 | | | <E-06 | 2.82E-07 |
| rs9850533 | 3 | | | <E-06 | 3.33E-07 |
| rs7314348 | 12 | | | <E-06 | 3.38E-07 |
| rs1940720 | 11 | NCAM1 | Schizophrenia (S2) | <E-06 | 3.50E-07 |
| rs2324317 | 11 | | | <E-06 | 3.59E-07 |
| rs1940718 | 11 | NCAM1 | Schizophrenia (S2) | <E-06 | 3.76E-07 |
| rs2298526 | 11 | NCAM1 | Schizophrenia (S2) | <E-06 | 3.76E-07 |
| rs8028364 | 15 | | | <E-06 | 3.78E-07 |
| rs10864120 | 1 | | | <E-06 | 4.27E-07 |
| rs4479020 | 11 | NCAM1 | Schizophrenia (S2) | <E-06 | 4.32E-07 |
| rs32215 | 5 | FBN2 | Congenital contractural arachnodactyly | <E-06 | 4.53E-07 |
| rs7121047 | 11 | NCAM1 | Schizophrenia (S2) | <E-06 | 4.98E-07 |
| rs1587327 | 11 | | | <E-06 | 4.99E-07 |
| rs11251291 | 10 | | | <E-06 | 6.19E-07 |
| rs961143 | 4 | | | <E-06 | 6.49E-07 |
| rs11251292 | 10 | | | <E-06 | 7.35E-07 |
| rs12626328 | 21 | | | <E-06 | 7.81E-07 |
| rs11724067 | 4 | | | <E-06 | 8.48E-07 |
| rs10937187 | 3 | MAGEF1 | | <E-06 | 8.51E-07 |
| rs7126748 | 11 | NCAM1 | Schizophrenia (S2) | <E-06 | 9.30E-07 |
| rs13233180 | 7 | SEPT14 | Glioblastoma | <E-06 | 9.66E-07 |
| rs4895650 | 6 | UTRN | Schizophrenia (S2) | <E-06 | 9.72E-07 |
| rs11710643 | 3 | | | <E-06 | 9.97E-07 |
| rs10751039 | 11 | | | <E-06 | 1.01E-06 |
| rs11251289 | 10 | | | <E-06 | 1.02E-06 |
| rs28579610 | 10 | | | <E-06 | 1.05E-06 |
| rs11251286 | 10 | | | <E-06 | 1.15E-06 |
| rs12359941 | 10 | | | <E-06 | 1.16E-06 |
| rs12817135 | 12 | | | <E-06 | 1.19E-06 |
| rs12817138 | 12 | | | <E-06 | 1.19E-06 |
| rs7105462 | 11 | NCAM1 | Schizophrenia (S2) | <E-06 | 1.21E-06 |
| rs1155447 | 10 | | | <E-06 | 1.32E-06 |
| rs12356215 | 10 | | | <E-06 | 1.38E-06 |
| rs7673072 | 4 | | | <E-06 | 1.41E-06 |
| rs11251283 | 10 | | | <E-06 | 1.41E-06 |
| rs2029604 | 10 | | | <E-06 | 1.41E-06 |
| rs4937870 | 11 | | | <E-06 | 1.49E-06 |
| rs7673034 | 4 | | | <E-06 | 1.50E-06 |
| rs17340760 | 12 | | | <E-06 | 1.51E-06 |
| rs10751040 | 11 | | | <E-06 | 1.59E-06 |
| rs58624222 | 1 | | | <E-06 | 1.60E-06 |
| rs4615154 | 4 | | | <E-06 | 1.61E-06 |
| rs593748 | 11 | | | <E-06 | 1.65E-06 |
| rs9733402 | 10 | | | <E-06 | 1.66E-06 |
| rs10027852 | 4 | | | <E-06 | 1.66E-06 |
| rs4403029 | 4 | | | <E-06 | 1.77E-06 |
| rs11532093 | 11 | | | <E-06 | 1.85E-06 |
| rs71661923 | 1 | | | <E-06 | 1.88E-06 |
| rs331090 | 5 | FBN2 | Congenital contractural | <E-06 | 1.92E-06 |
| | | | arachnodactyly | | |
| rs4034950 | 4 | | | <E-06 | 1.92E-06 |
| rs12679517 | 8 | | | <E-06 | 1.96E-06 |
| rs7115124 | 11 | | | <E-06 | 2.03E-06 |
| rs7119114 | 11 | | | <E-06 | 2.05E-06 |
| rs7305111 | 12 | | | <E-06 | 2.10E-06 |
| rs11712584 | 3 | MAGEF1 | | <E-06 | 2.21E-06 |
| rs138810467 | 12 | | | <E-06 | 2.23E-06 |
| rs10903785 | 10 | | | <E-06 | 2.33E-06 |
| rs10791113 | 11 | | | <E-06 | 2.33E-06 |
| rs11596010 | 10 | | | <E-06 | 2.35E-06 |
| rs4452753 | 7 | SEPT14 | Glioblastoma | <E-06 | 2.44E-06 |
| rs1117428 | 12 | | | <E-06 | 2.47E-06 |
| rs10243009 | 7 | SEPT14 | Glioblastoma | <E-06 | 2.55E-06 |
| rs6592883 | 11 | | | <E-06 | 2.65E-06 |
| rs4976939 | 8 | | | <E-06 | 2.67E-06 |
| rs10032875 | 4 | | | <E-06 | 2.67E-06 |
| rs4510809 | 7 | SEPT14 | Glioblastoma | <E-06 | 2.71E-06 |
| rs4439551 | 11 | NCAM1 | Schizophrenia (S2) | <E-06 | 2.82E-06 |
| rs1940726 | 11 | NCAM1 | Schizophrenia (S2) | <E-06 | 2.88E-06 |
| rs9302051 | 13 | ABCC4 | Prostatic Neoplasms | <E-06 | 2.94E-06 |
| rs4754522 | 11 | | | <E-06 | 2.96E-06 |
| rs30645 | 5 | ADAMTS19 | Premature Menopause | <E-06 | 3.03E-06 |
| rs4720832 | 7 | | | <E-06 | 3.04E-06 |
| rs847501 | 14 | | | <E-06 | 3.08E-06 |
| rs641472 | 1 | | | <E-06 | 3.11E-06 |
| rs512802 | 1 | | | <E-06 | 3.31E-06 |
| rs6972161 | 7 | SEPT14 | Glioblastoma | <E-06 | 3.32E-06 |
| rs11929982 | 4 | RELL1 | Liver Cirrhosis, Experimental | <E-06 | 3.50E-06 |
| rs2212328 | 11 | | | <E-06 | 3.63E-06 |
| rs11180229 | 12 | | | <E-06 | 3.65E-06 |
| rs10891492 | 11 | NCAM1 | Schizophrenia (S2) | <E-06 | 3.65E-06 |
| rs4476962 | 7 | SEPT14 | Glioblastoma | <E-06 | 3.72E-06 |
| rs9656760 | 8 | | | <E-06 | 3.72E-06 |
| rs228768 | 17 | HDAC5 | Mental Depression (S7) | <E-06 | 3.76E-06 |
| rs11715252 | 3 | | | <E-06 | 3.89E-06 |
| rs2155292 | 11 | NCAM1 | Schizophrenia (S2) | <E-06 | 3.92E-06 |
| rs3802847 | 11 | NCAM1 | Schizophrenia (S2) | <E-06 | 3.99E-06 |
| rs3924086 | 7 | | | <E-06 | 4.23E-06 |
| rs7942723 | 11 | NCAM1 | Schizophrenia (S2) | <E-06 | 4.26E-06 |
| rs7303492 | 12 | | | <E-06 | 4.39E-06 |
| rs10894306 | 11 | | | <E-06 | 4.47E-06 |
| rs331092 | 5 | FBN2 | Congenital contractural arachnodactyly | <E-06 | 4.53E-06 |
| rs1940713 | 11 | NCAM1 | Schizophrenia (S1, S2) | <E-06 | 4.57E-06 |
| rs1940714 | 11 | NCAM1 | Schizophrenia (S1, S2) | <E-06 | 4.57E-06 |
| rs4480572 | 11 | NCAM1 | Schizophrenia (S2) | <E-06 | 4.57E-06 |
| rs7935745 | 11 | NCAM1 | Schizophrenia (S2) | <E-06 | 4.57E-06 |
| rs34988996 | 10 | | | <E-06 | 4.62E-06 |
| rs61835737 | 10 | | | <E-06 | 4.62E-06 |
| rs7938812 | 11 | NCAM1 | Schizophrenia (S2) | <E-06 | 4.64E-06 |
| rs3802848 | 11 | NCAM1 | Schizophrenia (S2) | <E-06 | 4.70E-06 |
| rs10212515 | 3 | MAGEF1 | | <E-06 | 4.73E-06 |
| rs10927075 | 1 | AKT3 | Schizophrenia (S5) | <E-06 | 4.76E-06 |
| rs2052801 | 8 | | | <E-06 | 4.79E-06 |
| rs4589334 | 11 | NCAM1 | Schizophrenia (S2) | <E-06 | 4.80E-06 |
| rs1892981 | 11 | NCAM1 | Schizophrenia (S2) | <E-06 | 4.94E-06 |
| rs1940699 | 11 | NCAM1 | Schizophrenia (S2) | <E-06 | 4.99E-06 |
| rs34691721 | 8 | SCRIB | Neural Tube Defects (S4) | <E-06 | 5.01E-06 |
| rs2186874 | 11 | NCAM1 | Schizophrenia (S2) | <E-06 | 5.06E-06 |
| rs2155646 | 11 | NCAM1 | Schizophrenia (S2) | <E-06 | 5.10E-06 |
| rs7947502 | 11 | NCAM1 | Schizophrenia (S2) | <E-06 | 5.13E-06 |
| rs4294596 | 11 | NCAM1 | Schizophrenia (S2) | <E-06 | 5.23E-06 |
| rs7113099 | 11 | NCAM1 | Schizophrenia (S2) | <E-06 | 5.31E-06 |
| rs12011 | 3 | MAGEF1 | | <E-06 | 5.52E-06 |
| rs1940697 | 11 | NCAM1 | Schizophrenia (S2) | <E-06 | 5.82E-06 |
| rs7128314 | 11 | NCAM1 | Schizophrenia (S2) | <E-06 | 5.90E-06 |
| rs2212449 | 11 | NCAM1 | Schizophrenia (S2) | <E-06 | 5.94E-06 |
| rs695134 | 11 | IGSF9B | Schizophrenia (S2) | <E-06 | 5.94E-06 |
| rs331089 | 5 | FBN2 | Congenital contractural arachnodactyly | <E-06 | 6.03E-06 |
| rs723599 | 11 | NCAM1 | Schizophrenia (S2) | <E-06 | 6.12E-06 |
| rs4820386 | 22 | CAC1I | | <E-06 | 6.13E-06 |
| rs999851 | 11 | NCAM1 | Schizophrenia (S2) | <E-06 | 6.20E-06 |
| rs7113596 | 11 | NCAM1 | Schizophrenia (S2) | <E-06 | 6.30E-06 |
| rs10891487 | 11 | NCAM1 | Schizophrenia (S2) | <E-06 | 6.44E-06 |
| rs507378 | 11 | IGSF9B | Schizophrenia (S2) | <E-06 | 6.59E-06 |
| rs10750019 | 11 | NCAM1 | Schizophrenia (S2) | <E-06 | 6.67E-06 |
| rs7127712 | 11 | NCAM1 | Schizophrenia (S2) | <E-06 | 6.73E-06 |
| rs7127930 | 11 | NCAM1 | Schizophrenia (S2) | <E-06 | 6.73E-06 |
| rs1940733 | 11 | NCAM1 | Schizophrenia (S2) | <E-06 | 6.74E-06 |
| rs4821910 | 22 | CAC1I | | <E-06 | 6.81E-06 |
| rs2186707 | 11 | NCAM1 | Schizophrenia (S2) | <E-06 | 6.87E-06 |
| rs13147397 | 4 | | | <E-06 | 6.91E-06 |
| rs9919620 | 11 | NCAM1 | Schizophrenia (S2) | <E-06 | 6.92E-06 |
| rs10750021 | 11 | NCAM1 | Schizophrenia (S2) | <E-06 | 7.15E-06 |
| rs10092551 | 8 | SCRIB | Neural Tube Defects | <E-06 | 7.20E-06 |
| rs1940725 | 11 | NCAM1 | Schizophrenia (S2) | <E-06 | 7.32E-06 |
| rs1940727 | 11 | NCAM1 | Schizophrenia (S2) | <E-06 | 7.32E-06 |
| rs1954826 | 11 | NCAM1 | Schizophrenia (S2) | <E-06 | 7.32E-06 |
| rs1940702 | 11 | NCAM1 | Schizophrenia (S1, S2) | <E-06 | 7.93E-06 |
| rs4144892 | 11 | NCAM1 | Schizophrenia (S2) | <E-06 | 8.14E-06 |
| rs1940716 | 11 | NCAM1 | Schizophrenia (S2, S6) | <E-06 | 8.20E-06 |
| rs7950836 | 11 | NCAM1 | Schizophrenia (S2) | <E-06 | 8.20E-06 |
| rs9919670 | 11 | NCAM1 | Schizophrenia (S11) | <E-06 | 8.54E-06 |
| rs1940701 | 11 | NCAM1 | Schizophrenia (S2, S5) | <E-06 | 8.82E-06 |
| rs12739344 | 1 | AKT3 | Schizophrenia (S5) | <E-06 | 8.95E-06 |
| rs7108081 | 11 | NCAM1 | Schizophrenia (S9) | <E-06 | 9.03E-06 |
| rs4255098 | 8 | | | <E-06 | 9.17E-06 |
| rs1911723 | 12 | | | <E-06 | 9.78E-06 |
| rs12538049 | 7 | SEPT14 | Glioblastoma | <E-06 | 9.96E-06 |
| rs8097665 | 18 | DOK6 | Tobacco Use Disorder | <E-06 | 1.03E-05 |
| rs4733144 | 8 | | | <E-06 | 1.05E-05 |
| rs876983 | 8 | PSD3 | Schizophrenia (S1) | <E-06 | 1.42E-05 |
| rs11993154 | 8 | SCRIB | Neural Tube Defects (S4) | <E-06 | 1.51E-05 |
| rs1911724 | 12 | | | <E-06 | 1.56E-05 |
| rs16875703 | 8 | | | <E-06 | 1.69E-05 |
| rs4947524 | 7 | | | <E-06 | 1.78E-05 |
| rs2039461 | 9 | | | <E-06 | 1.96E-05 |
| rs4869972 | 6 | MTHFD1L | Alzheimer's Disease (S8) | <E-06 | 1.98E-05 |
| rs12419623 | 11 | RS2 | | <E-06 | 2.22E-05 |
| rs7035838 | 9 | | | <E-06 | 2.23E-05 |
| rs714031 | 22 | CAC1I | | <E-06 | 2.23E-05 |
| rs10075211 | 5 | HTR4 | Schizophrenia (S1) | <E-06 | 2.24E-05 |
| rs2409138 | 21 | | | <E-06 | 2.34E-05 |
| rs3812442 | 8 | TSTA3 | Malignt neoplasm of breast | <E-06 | 2.50E-05 |
| rs6468144 | 8 | | | <E-06 | 2.60E-05 |
| rs6582447 | 12 | | | <E-06 | 3.14E-05 |
| rs55648724 | 6 | | | <E-06 | 3.33E-05 |
| rs7775684 | 6 | | | <E-06 | 3.34E-05 |
| rs7385206 | 7 | | | <E-06 | 3.56E-05 |
| rs5757766 | 22 | CAC1I | | <E-06 | 3.63E-05 |
| rs725515 | 16 | CDH13 | Mental Depression (S10) | <E-06 | 3.80E-05 |
| rs7014454 | 8 | | | <E-06 | 3.90E-05 |
| rs114180690 | 6 | | | <E-06 | 3.95E-05 |
| rs866154 | 10 | | | <E-06 | 4.09E-05 |
| rs116414669 | 6 | | | <E-06 | 5.63E-05 |
| rs114954038 | 6 | | | <E-06 | 6.24E-05 |
| rs34259118 | 9 | BNC2 | Craniofacial Abnormalities | <E-06 | 8.50E-05 |
| rs7283105 | 21 | | | <E-06 | 9.01E-05 |
| rs10986439 | 9 | GABBR2 | Major Depressive Disorder (S3) | <E-06 | 0.00045792 |
| rs181555294 | 6 | | | <E-06 | 0.00087933 |
| rs4531108 | 9 | | | <E-06 | 0.00186101 |

Table S2. Power to detect association between SNPs and disease

| Sample Size | 500 | 1000 | 2000 | 5000 |
|---|---|---|---|---|
| 0.05 | 0.999 | 1 | 1 | 1 |
| 0.01 | 0.992 | 0.992 | 0.993 | 0.992 |

Table S3. Power to detect association between SNP1 and Disease.

| Sample Sizes | 500 | 1000 | 2000 | 5000 |
|---|---|---|---|---|
| 0.05 | 0.999 | 1 | 1 | 0.999 |
| 0.01 | 0.992 | 0.992 | 0.993 | 0.992 |

Table S4. Power to detect association between SNP1 and disease.

| Sample Sizes | 500 | 1000 | 2000 | 5000 |
|---|---|---|---|---|
| 0.05 | 0.684 | 0.888 | 0.949 | 0.949 |
| 0.01 | 0.418 | 0.701 | 0.936 | 0.948 |